\pdfoutput=1
\documentclass[11pt,twoside,a4paper,cmspaper,final,collab]{cms-tdr}
\def\svnVersion{181ffa7-D}\def\svnDate{2021/05/18}\def\cmsCernNoTag{CERN-EP-2020-247}\def\cmsCernDate{\today}\def\cmsMessage{"Published in the Journal of High Energy Physics as \href{http://dx.doi.org/10.1007/JHEP05(2021)116}{\doi{10.1007/JHEP05(2021)116}.}"}
\begin{document}\cmsNoteHeader{HIN-19-013}

\newcommand{\pp}{\ensuremath{\Pp\Pp}\xspace}
\newcommand{\ptSuper}[1]{\ensuremath{\pt^{\mathrm{#1}}}\xspace}
\newcommand{\ptText}[1]{\ensuremath{\pt^{\text{#1}}}\xspace}
\newcommand{\ptCh}{\ptSuper{ch}}
\newcommand{\ptLead}{\ptText{leading}}
\newcommand{\ptSub}{\ptText{subleading}}
\newcommand{\rhoSub}[1]{\ensuremath{\rho(\Delta r)_{\mathrm{#1}}}\xspace}
\newcommand{\rhoPbPb}{\rhoSub{PbPb}}
\newcommand{\rhoPp}{\rhoSub{\pp}}
\newcommand{\xj}{\ensuremath{x_{j}}\xspace}
\newcommand{\NText}[1]{\ensuremath{N_{\text{#1}}}\xspace}
\newcommand{\Ndijet}{\NText{dijet}}

\newlength\cmsTabSkip\setlength{\cmsTabSkip}{1ex}

\cmsNoteHeader{HIN-19-013} 

\title{In-medium modification of dijets in PbPb collisions at \texorpdfstring{$\sqrtsNN = 5.02\TeV$}{sqrt(sNN) = 5.02 TeV}}

\date{\today}

\abstract{
   Modifications to the distribution of charged particles with respect to high transverse momentum (\pt) jets passing through a quark-gluon plasma are explored using the CMS detector. Back-to-back dijets are analyzed in lead-lead and proton-proton collisions at $\sqrtsNN = 5.02\TeV$ via correlations of charged particles in bins of relative pseudorapidity and angular distance from the leading and subleading jet axes. In comparing the lead-lead and proton-proton collision results, modifications to the charged-particle relative distance distribution and to the momentum distributions around the jet axis are found to depend on the dijet momentum balance \xj, which is the ratio between the subleading and leading jet \pt. For events with $\xj \approx 1$, these modifications are observed for both the leading and subleading jets. However, while subleading jets show significant modifications for events with a larger dijet momentum imbalance, much smaller modifications are found for the leading jets in these events.
}

\hypersetup{
pdfauthor={CMS Collaboration},
pdftitle={In-medium modification of dijets in PbPb collisions at sqrt(s[NN]) = 5.02 TeV},
pdfsubject={CMS},
pdfkeywords={CMS, physics, jets, jet shape, dijet, correlations}}

\maketitle

\section{Introduction}
\label{sec:intro}

In relativistic heavy ion collisions, high transverse momentum (\pt) jets originate from partons that have undergone a hard scattering and may be used to probe the properties of the quark-gluon plasma (QGP) created in such collisions~\cite{heavyIonReview}. One phenomenon related to the properties of the QGP is parton energy loss~\cite{BjorkenQGP}, also known as jet quenching, which was first observed at the BNL RHIC~\cite{Adams:2006yt, Adare:2007vu} and subsequently at the CERN LHC~\cite{Aad:2010bu, Chatrchyan:2011sx, centralityDetermination, Adam:2015ewa}.  Jet quenching is seen as a suppression of high-\pt leading charged-particle and jet yields in heavy ion collisions relative to a proton-proton (\pp) reference~\cite{qgpBrahms, qgpPhobos, qgpStar, qgpPhenix}. Using data collected at the LHC, studies have shown that the jet structure is also modified by the medium, as observed with measurements of the fragmentation functions~\cite{CMS:2012vba, Aad:2014wha}, and by the distribution of charged-particle \pt as a function of the radial distance from the jet axis~\cite{Chatrchyan:2013kwa}. These modifications are found to extend to large distances in relative pseudorapidity ($\Delta\eta$) and relative azimuth ($\Delta\varphi$) with respect to the jet axis~\cite{mpt, Khachatryan:2016erx, lowerEnergyDijet, inclusiveJetShape}. Various theoretical models have  attempted to account for these modifications~\cite{Apolinario:2012cg, Ayala:2015jaa, Blaizot:2014ula, Escobedo:2016jbm, Casalderrey-Solana:2016jvj, Tachibana:2017syd} and while most models reproduce the modifications close to the jet axis, the large modifications observed far from the jet axis $\Delta r = \sqrt{\smash[b]{(\Delta\varphi)^{2} + (\Delta\eta)^{2}}} > 0.5$ are not yet well modeled. 

The analysis presented in this paper uses LHC data collected by the CMS experiment at a collision energy of $\sqrtsNN = 5.02\TeV$ in 2017 and 2018, corresponding to integrated luminosities of 320\pbinv for \pp and 1.7\nbinv for lead-lead (PbPb) collisions, respectively. Events are selected with nearly back-to-back ($\Delta\varphi > 5 \pi / 6$), high-\pt leading and subleading jet pairs. Correlations in relative pseudorapidity are measured for charged particles with respect to both the leading and subleading jet axes, where the relative azimuthal angle between the jet axes and charged particles is restricted to $\abs{\Delta\varphi} < 1.0$.  The ``jet shape," which is the distribution of charged-particle transverse momentum (\ptCh) as a function of the distance from the jet axis $\Delta r$, is also studied.  Results are presented differentially as functions of PbPb collision centrality (\ie, the degree of overlap of the colliding nuclei, with head-on collisions defined as ``most central")~\cite{centralityDetermination}, \ptCh, and dijet momentum balance $\xj = \ptSub / \ptLead$. Compared to previous jet shape analyses in Refs.~\cite{lowerEnergyDijet, inclusiveJetShape}, the large PbPb data sample recorded in 2018 allows for differentiation in \xj for leading and subleading jet shapes extended to large distances from the jet axis.

\section{The CMS experiment}
\label{sec:cms}

The central feature of the CMS apparatus is a superconducting solenoid of 6\unit{m} internal diameter, providing a magnetic field of 3.8\unit{T}. Within the solenoid volume are a silicon pixel and strip tracker, a lead tungstate crystal electromagnetic calorimeter (ECAL), and a brass and scintillator hadron calorimeter (HCAL), each composed of barrel and two endcap sections. Two hadronic forward (HF) steel and quartz-fiber calorimeters complement the barrel and endcap detectors, extending the calorimeter from the range $\abs{\eta} < 3.0$ provided by the barrel and endcap out to $\abs{\eta} < 5.2$. The HF calorimeters are segmented to form $0.175{\times}0.175$ ($\Delta\eta{\times}\Delta\varphi$) towers. The sum of the transverse energies detected in the HF detectors $(3.0 < \abs{\eta} < 5.2)$ is used to define the event centrality in PbPb events and to divide the event sample into centrality classes, each representing a percentage of the total nucleus-nucleus hadronic interaction cross section.  A detailed description of the centrality determination can be found in Ref.~\cite{centralityDetermination}.

Jets used in this analysis are reconstructed within the range $\abs{\eta} < 1.6$. In the region $\abs{\eta} < 1.74$, the HCAL cells have widths of 0.087 in both $\eta$ and $\varphi$ and thus provide high granularity. Within the central barrel region of $\abs{\eta} < 1.48$, the HCAL cells map onto $5{\times}5$ ECAL crystal arrays to form calorimeter towers projecting radially outwards from the nominal interaction point. Within each tower, the energy deposits in ECAL and HCAL cells are summed to define the calorimeter tower energy. 

The CMS silicon tracker measures charged-particle tracks within $\abs{\eta} < 2.5$.  It consists of 1856 silicon pixel and $15\,148$ silicon strip detector modules. Muons are measured in the pseudorapidity range $\abs{\eta} < 2.4$, with detection planes made using three technologies: drift tubes, cathode strip chambers, and resistive plate chambers. 

Events of interest are selected using a two-tiered trigger system. The first level (L1), composed of custom hardware processors, uses information from the calorimeters and muon detectors to select events at a rate of around 100\unit{kHz}~\cite{Sirunyan:2020zal}. The second level, known as the high-level trigger (HLT), consists of a farm of processors running a version of the full event reconstruction software optimized for fast processing, and reduces the event rate to around 1\unit{kHz} before data storage~\cite{Khachatryan:2016bia}. 

A detailed description of the CMS detector, together with a definition of the coordinate system used and the relevant kinematic variables, can be found in Ref.~\cite{CMSexperiment}.

\section{Event selection}
\label{sec:eventSelection}

The \pp and PbPb data are selected with a calorimeter-based HLT trigger that uses the anti-\kt jet clustering algorithm with a distance parameter of $R = 0.4$~\cite{antiKt}. The trigger requires events to contain at least one jet with $\pt > 80$ and $> 100\GeV$ for \pp and PbPb collisions, respectively. For PbPb collisions, the underlying event contribution is subtracted from the jet \pt using an iterative method~\cite{puSubtraction} before comparing to the threshold. The data sample selected by this trigger is referred to as ``jet-triggered." For the PbPb event selection, a minimum bias triggered~\cite{Chatrchyan:2011pb} sample is also used in the analysis. 

To reduce contamination from noncollision events, including calorimeter noise and beam-gas collisions, vertex and noise filters are applied offline to both the \pp and PbPb data, following previous analyses of lower-energy data~\cite{Chatrchyan:2011sx, centralityDetermination}. These filters include requirements that at least three HF towers on either side of the interaction point have a tower energy above 3\GeV  and that the vertex position along the beam line lies within 15\unit{cm} of the nominal interaction point. The average pileup (the number of collisions per beam bunch crossing) is about 2 in \pp, and close to 1 in PbPb collisions. Because of the low pileup in the data samples, no pileup corrections are necessary.

Monte Carlo simulated event samples are used to evaluate the performance of the event reconstruction, particularly the track reconstruction efficiency, and the jet energy response and resolution. The hard scattering, parton shower, and fragmentation of the partons are modeled using the \PYTHIA~8 event generator with tune CP5~\cite{pythia82, pythiaCP5tune}. The specific \PYTHIA version used is 8.226 and the parton distribution function set is NNPDF3.1 at next-to-next-to-leading order~\cite{nnpdf31}. The CMS detector response is simulated using the \GEANTfour toolkit~\cite{geant4}. The soft underlying event for the PbPb collisions is simulated by the \HYDJET 1.9 event generator~\cite{hydjet}. The energy density in the \HYDJET simulation is tuned to match the data by shifting the centrality binning in standard \HYDJET simulation by 5 percentage points upwards. This tuning is based on a random cone study, where rigid cones with a radius 0.4 are placed in random directions in events, and the energy densities in the cones are determined by summing the particle transverse momenta within the cone. The above tuning provides the best agreement in random cone energy densities between data and \HYDJET. To simulate jet events in PbPb collisions, \PYTHIA~8 generated hard events are embedded into soft \HYDJET events. This sample is denoted as {\PYTHIA}+\HYDJET.

In PbPb collisions, jets are produced more frequently in central events than in noncentral events because of the large number of binary collisions per nuclear interaction. A centrality-based reweighting is applied to the {\PYTHIA}+\HYDJET sample in order to match the centrality distribution of the jet-triggered PbPb data. An additional reweighting procedure is performed to match the simulated vertex distributions to data for both the \pp and PbPb samples.

The event selection works as follows. First, the two jets with highest \pt are located in the range of $\abs{\eta} < 2$. The highest \pt jet is called the leading jet and it is required to pass a \pt cut of $p_{\mathrm{T},1} > 120\GeV$. The second highest \pt jet is termed the subleading jet and for it a cut $p_{\mathrm{T},2} > 50\GeV$ is applied. Then, the back-to-back requirement, a condition demanding that the jet separation in $\varphi$ obeys $\Delta\varphi_{1,2} > 5 \pi / 6$ is enforced. Finally, both jets are required to fall within $\abs{\eta} < 1.6$ to ensure the most stable jet reconstruction performance and to allow for good tracker acceptance on both sides of the jets. There is no veto for additional jets in the event. The events containing such pairs of back-to-back jets are referred as dijet events for the remainder of the paper.

\section{Jet and track reconstruction}
\label{sec:jetTrackReconstruction}

For this analysis, jets in both \pp and PbPb collisions are reconstructed using the anti-\kt algorithm with a distance parameter $R = 0.4$, as implemented in the \textsc{FastJet} framework~\cite{fastJet}. A particle flow (PF) algorithm using an optimized combination of information from various elements of the CMS detector is used to reconstruct leptons, photons, and charged and neutral hadrons~\cite{particleFlow}. These PF candidates are employed to reconstruct the jets used in this analysis. The jets are first clustered using E-scheme clustering~\cite{fastJet} and the jet axis is recalculated with the winner-take-all algorithm~\cite{wta1, wta2} using anti-\kt and the same constituents. In the E-scheme clustering, particle pairs are iteratively combined to form pseudo-jets (an object that is a combination of particles or other pseudo-jets), with the direction of the new pseudo-jet given by the sum of the four-momenta of the particles. In the winner-take-all scheme, the direction of the new pseudo-jet in each iteration is aligned with the direction of the particle or pseudo-jet with higher \pt. It follows that the E-scheme axis for the final jet is given by the sum of pseudo-jet four-momenta, while the winner-take-all axis is determined by the axis of the hardest pseudo-jet. The winner-take-all axis is preferred in this analysis over the default E-scheme axis because of an artificial feature created by the E-scheme axis that results in a strong depletion of particles just outside of the jet cone radius. Not having to correct for this feature leads to smaller uncertainties.

In order to subtract the soft underlying event contribution to the jet energy in PbPb collisions, a constituent subtraction method~\cite{constituentSubtraction} is employed. This involves a particle-by-particle approach that corrects jet constituents based on the local average underlying event density. The energy density is estimated using an event-by-event, iterative algorithm~\cite{puSubtraction} that finds the mean value, $\langle E_{\mathrm{PF}}\rangle$, and dispersion, $\sigma(E_{\mathrm{PF}})$, of the energies from the PF candidates in $\eta$ strips~\cite{Chatrchyan:2011sx,centralityDetermination}. In \pp collisions, where the underlying event level is negligible, jets are reconstructed without underlying event subtraction.

The track reconstruction used in \pp and PbPb collisions is described in Ref.~\cite{trackResolution}. For the charged-particle tracks used in jet to charged-particle correlations, it is required that the relative \pt uncertainty of each track is less than 10\%. In PbPb collisions, tracks must also have at least 11 hits in the tracker layers and satisfy a stringent fit quality requirement, specifically that the $\chi^{2}$, divided by the product of the number of fit degrees of freedom and the number of tracker layers hit, be less than 0.18. Furthermore, it is required that the significance of the distance of closest approach of a charged-particle track to at least one primary vertex in the event is less than 3 standard deviations. This is done to decrease the likelihood of counting nonprimary charged particles originating from secondary decay products, and is applied for both \pp and PbPb collisions. Finally, for PbPb collisions a selection based on the relationship of a track to the calorimeter energy deposits along its trajectory is applied in order to reduce the contribution of misreconstructed tracks with very high \pt. Tracks with $\pt > 20\GeV$ are required to have an associated energy deposit of at least half of their momentum in the CMS calorimeters. Corrections for tracking efficiency, detector acceptance, and misreconstruction rate are obtained and applied following the procedure of Ref.~\cite{Khachatryan:2016odn}. 

\section{Jet to charged-particle angular correlations}
\label{sec:analysisMethods}

\linespread{1.01}
\selectfont

Correlations between reconstructed jets and charged-particle tracks are studied by forming a two-dimensional distribution of $\Delta\eta$ and $\Delta\varphi$ of the charged particles relative to the jet axis.  Events in PbPb collisions are divided into four centrality intervals, 0--10, 10--30, 30--50, and 50--90\%, based on the total energy collected in the HF calorimeter~\cite{centralityDetermination}. The events are also binned by the charged-particle \pt with bin boundaries of 0.7, 1, 2, 3, 4, 8, 12, and 300\GeV and in the dijet momentum balance \xj with bin boundaries of 0, 0.6, 0.8, and 1. The effects of jet energy resolution to the \xj binning are explored in Appendix~\ref{app:Matrices}. The two-dimensional correlation histograms are filled by correlating all charged particles in the event with the leading jet and the subleading jet, separately. For the jet shape measurement, each entry is weighted by the charged particle \pt value. The histograms are then normalized by the number of dijets. The numbers of dijets in the data samples for each \xj and centrality bin are summarized in Table~\ref{tab:xjbinning}.

\begin{table}
\centering{
  \topcaption{The total number of events for \pp and for different PbPb centrality bins are shown in the top row. The other rows show the percentage of all events that falls within a given \xj bin.}
  \label{tab:xjbinning} 
  \begin{tabular}{cccccc}
    \xj &  0--10\%  & 10--30\% & 30--50\% & 50--90\% & \pp   \\
    \hline
    $0 < \xj < 1$       & $5.8 \times 10^{5} $  & $6.6 \times 10^{5}$ & $2.6 \times 10^{5}$ & $0.78 \times 10^{5}$ & $110 \times 10^{5}$ \\[\cmsTabSkip]
    $0 < \xj < 0.6$     & 40\% & 35\% & 29\% & 26\% & 24\% \\
    $0.6 < \xj < 0.8$   & 33\% & 34\% & 36\% & 35\% & 36\% \\
    $0.8 < \xj < 1$     & 27\% & 31\% & 35\% & 39\% & 40\% \\
  \end{tabular}
}
\end{table}

Since the detector has limited acceptance in $\eta$, it is more probable to find jet-charged-particle pairs with small rather than large $\Delta\eta$ values. Thus, the raw correlations have a shape where the charged-particle yield strongly decreases toward large $\Delta\eta$. A mixed-event method is employed to correct for these detector acceptance effects. In this method, the jet and charged particles from different events are mixed to ensure that no physical correlations exist in the resulting distribution. What is left is the structure arising from the detector acceptance. When the mixed events are constructed, it is required that the primary vertex positions along the beam axis match within 0.5\unit{cm} and that the centrality values match within 0.5 percentage points between the two events. For \pp collisions, a jet-triggered sample is used for the mixed events, while for PbPb collisions a minimum bias sample is used to properly capture the long range correlations. This procedure is similar to that used in previous analyses~\cite{lowerEnergyDijet, inclusiveJetShape}. Denoting the number of dijets satisfying the selection criteria as \Ndijet, the per-dijet associated charged-particle yield corrected for the acceptance effects is given by
\begin{linenomath}
\begin{equation}
  \frac{1}{\Ndijet} \frac{\rd^{2}N}{\rd\Delta\eta\rd\Delta\varphi} = \frac{ME(0,0)}{ME(\Delta\eta,\Delta\varphi)} S(\Delta\eta,\Delta\varphi),
  \label{eq:acceptanceCorrection}
\end{equation}
\end{linenomath}
where $N$ is the number of jet-particle pairs, the signal pair distribution $S(\Delta\eta,\Delta\varphi)$ represents the per-dijet normalized yield of jet-particle pairs from the same event,
\begin{linenomath}
\begin{equation}
  S(\Delta\eta,\Delta\varphi) = \frac{1}{\Ndijet} \frac{\rd^{2}N^{\text{same}}}{\rd\Delta\eta\rd\Delta\varphi},
  \label{eq:sameEvent}
\end{equation}
\end{linenomath}
and the mixed-event pair distribution $ME(\Delta\eta,\Delta\varphi)$ is given by
\begin{linenomath}
\begin{equation}
  ME(\Delta\eta,\Delta\varphi) = \frac{\rd^{2}N^{\text{mixed}}}{\rd\Delta\eta\rd\Delta\varphi}.
  \label{eq:mixedEvent}
\end{equation}
\end{linenomath}
The ratio $ME(0,0) / ME(\Delta\eta,\Delta\varphi)$ is the normalized correction factor. The maximum of the mixed event distribution can be found at $(0,0)$ as no pairs with $\Delta\eta = 0$ are lost as a consequence of acceptance effects.

The acceptance-corrected distribution has contributions from several sources. The jet correlation shows up as a Gaussian peak around $(\Delta\eta,\Delta\varphi) = (0,0)$ together with a peak elongated in $\Delta\eta$ around $\Delta\varphi = \pi$. While the underlying event contribution to jet energy and momentum is corrected for as previously discussed in Section~\ref{sec:jetTrackReconstruction}, the underlying event particles paired with the jet remain within the acceptance-corrected distribution, and thus have to be removed. To model this background, the $\Delta\varphi$ distribution is averaged over the region $1.5 < \abs{\Delta\eta} < 2.5$ on the near side ( $\abs{\Delta\varphi} < \pi/2$) of the jet. The same background region criteria are used for correlations of charged particles with both leading and subleading jets and the backgrounds are combined to cover the full $\Delta\varphi$ range. This procedure is applied to avoid an ``eta swing" effect. Momentum conservation dictates that in a statistical ensemble of dijets, the two jets must be approximately back-to-back in $\Delta\varphi$, but no such requirement exists for the $\Delta\eta$ separation. Thus, the away side $(\abs{\Delta\varphi} > \pi/2 )$ jet peak is prolonged in $\Delta\eta$ in jet to charged-particle correlation distributions. To avoid mixing the jet signal with the background, the away side region needs to be avoided in the background estimation. The background determined using the respective near-side components is propagated to the full $(\Delta\eta,\Delta\varphi)$ plane and subtracted from the acceptance-corrected distribution to obtain the jet signal.

Finally, simulation-based corrections are applied to account for a bias toward selecting jets with a harder constituent \pt spectrum (affecting PbPb and \pp events similarly) and a bias toward selecting jets that are affected by upward fluctuations in the soft underlying event yield (relevant for PbPb events only). Jets with a harder constituent \pt spectrum are more likely to be successfully reconstructed than jets with a softer \pt spectrum because the calorimeter response does not scale linearly with the incident particle energy, resulting in a bias toward the selection of jets with fewer associated particles. A residual correction for this bias is derived and applied following the method described in Refs.~\cite{mpt, Khachatryan:2016erx, lowerEnergyDijet}, by comparing per-jet yields of generated particles correlated to reconstructed jets relative to those correlated to generated jets. This correction is derived using a \PYTHIA~8 simulation for \pp events and considering only generated particles coming from the embedded \PYTHIA hard process in {\PYTHIA}+\HYDJET simulation for PbPb events.

For PbPb events, there is an additional jet reconstruction bias toward the selection of jets that are produced in the vicinity of upward fluctuations in the underlying event background. The underlying event fluctuations together with a steeply falling jet \pt spectrum produce a resolution effect on the jet energy scale. This causes more jets associated with upward fluctuations than with downward fluctuations of the underlying event to pass the analysis threshold. To account for this bias, a similar procedure to that outlined in Refs.~\cite{Khachatryan:2016erx, lowerEnergyDijet, Chatrchyan:2012gw} is followed. Correlations in the {\PYTHIA}+\HYDJET sample between reconstructed jets and generated particles are constructed using only particles from the \HYDJET underlying event, excluding those from the embedded hard process. This gives an estimate of the underlying event yield on top of which the jet correlations are found in the data. To reduce the fluctuations in the generated sample, the obtained distribution is symmetrized in $\Delta\eta$ and $\Delta\varphi$, before being applied as a correction to the PbPb data.

\linespread{1.00}
\selectfont

\section{Systematic uncertainties}
\label{sec:systematics}

The following sources of uncertainty are considered in this analysis:
\begin{itemize}
  \item \emph{Underlying event fluctuation bias}. This source accounts for uncertainties in correcting for the jet reconstruction bias resulting from upwards fluctuations of the underlying event background. Three different causes are considered. The first is a potential difference between the quark and gluon jet fraction in simulation and in data. The potential difference is estimated to be less than 25\% using a template fit to the multiplicity distribution of PF candidates within the jet cone in the data. Then the uncertainty is estimated by varying the quark/gluon jet fraction in simulation by this amount. The second source considers the difference in the underlying event energy density between simulation and data, and is estimated by random cone studies. The uncertainty is determined by varying the centrality binning of the \HYDJET simulation by 1 percentage point around the best match to data, which is the range giving a reasonable agreement. Finally, there is the uncertainty in the underlying event level for the simulation from which the underlying event fluctuation bias correction is derived. This uncertainty is obtained in a similar manner as for data, as discussed below in the ``underlying event subtraction" bullet point. 
  \item \emph{Jet fragmentation bias}. Most of the detector and resolution effects contributing to the jet fragmentation bias uncertainty arise from the uncertainty in the ratios of quark and gluon jets between data and simulation, as discussed for the underlying event fluctuation bias above. Deriving the corrections separately for quark and gluon jets and varying their relative contribution within the estimated difference between data and Monte Carlo, a 10\% variation in the fragmentation bias correction is observed. This difference is assigned as a systematic uncertainty.
  \item \emph{Jet energy scale}. These uncertainties are estimated by varying the jet energy corrections within their uncertainties and seeing how these changes affect the final correlations. The uncertainties of the jet energy corrections are discussed in detail in Ref.~\cite{jetEnergyScale}.
  \item \emph{Jet energy resolution}. This uncertainty is estimated by adding a Gaussian spread to the nominal jet energies such that the jet energy resolution estimated from the simulation is worsened by 20\% and comparing the obtained results to the nominal ones. The value of 20\% comes from the maximal estimated difference in the jet energy resolution between data and simulation.
  \item T\emph{rigger bias}. The calorimeter-based trigger with a threshold of 100\GeV is not fully efficient for the PbPb collisions. To see if this has an effect on the final results, the analysis was repeated requiring a prescaled trigger with a threshold of 80\GeV and the results with this trigger were compared to the nominal ones. For the leading jet shapes, it was found that there is a 2\% yield difference in the bin closest to the jet axis, while for the other bins and for the subleading jets the difference is negligible. The 2\% difference is therefore applied as a systematic uncertainty on the first bin. The trigger used for the \pp collisions is fully efficient and thus has no trigger bias uncertainty.
  \item \emph{Tracking efficiency}. Two sources of uncertainty are considered. First, there are possible track reconstruction differences in data and simulation. Following the method in Ref.~\cite{Khachatryan:2016odn}, it is estimated that the uncertainty in the track reconstruction efficiency and misidentification rate corrections is 5\% for PbPb and 2.4\% for \pp collisions. Second, the ratio of corrected reconstructed yields to generator level yields is studied for \PYTHIA~8 and {\PYTHIA}+\HYDJET simulations. It is found that the reconstructed and generated yields match within 3\% in PbPb and 1\% in \pp collisions, and these numbers are used as a systematic uncertainty from this source. An extra uncertainty is added to tracks close to the jet axis, as in the high multiplicity environment around the jet, the tracking efficiency is found to be 1--2\% worse than far away from the jets.
  \item \emph{Acceptance correction}. In an ideal case, the mixed-event corrected jet to charged-particle angular distribution contains a jet peak and an uncorrelated underlying event background. Since jet correlations are small angle correlations, the underlying event is expected to dominate far from the jet axis (sideband), and thus the distribution at $\abs{\eta} > 1.5$ should be uniform. To evaluate possible deviations from the uniformity, a constant function is separately fit to each sideband region of the acceptance-corrected $\Delta\eta$ distribution ($-2.5<\Delta\eta<-1.5$ and $1.5<\Delta\eta<2.5$). The difference between the average yield in the positive and negative sides determined this way is assigned as a systematic uncertainty.
  \item \emph{Underlying event subtraction}. Uncertainties resulting from the underlying event subtraction are determined by considering the two parts of the sideband region, $1.5<\abs{\Delta\eta}<2.0$ and $2.0<\abs{\Delta\eta}<2.5$, after the underlying event subtraction. The average yield is calculated in each of the regions and the larger deviation from zero is assigned as a systematic uncertainty.
\end{itemize}

The total systematic uncertainties are obtained by adding all the individual components together in quadrature. For the leading and  subleading jet correlations, the uncertainties integrated over \xj and $\Delta r$ are listed in Tables~\ref{tab:leadingJetSystematics} and~\ref{tab:subleadingJetSystematics}, respectively.

\begin{table}[htbp]
  \centering{
    \topcaption{Systematic uncertainties for the leading jet shape components, integrated over \xj, and $\Delta r$, and shown for \pp and centrality-binned PbPb collisions. The ranges correspond to the \pt dependence of the uncertainty. If some \pt bins have an uncertainty smaller than 0.5\%, the range is presented with a ``$<$" symbol and the upper bound.}
    \label{tab:leadingJetSystematics}
    \begin{tabular}{cccccc}
      Source & 0--10\% &  10--30\% & 30--50\%& 50--90\% & \pp\\ 
      \hline
      Underlying event fluctuation bias & $<$4\% & $<$3\% & $<$2\% & $<$1\% & \NA \\
      Jet fragmentation bias  & $<$3\% & $<$2\% & $<$2\% & $<$2\% & $<$1\% \\
      Residual jet energy scale & 3--7\% & 3--8\% & 3--8\% & 3--9\% & 1--9\% \\
      Jet energy resolution & 1--3\% & 1--3\% & 1--3\% & 1--3\% & $<$2\% \\
      Trigger bias & $<$2\% & $<$2\% & $<$2\% & $<$2\% & \NA \\
      Tracking efficiency & 6--8\% & 6--8\% & 6--7\% & 6--7\% & 3\% \\
      Acceptance correction & $<$3\% & $<$3\% & $<$3\% & $<$3\% & $<$1\% \\
      Underlying event subtraction & $<$4\% & $<$3\% & $<$3\% & $<$3\% & $<$2\% \\[\cmsTabSkip]
      Total & 8--14\% & 8--12\% & 8--12\% & 8--11\% & 3--10\% \\
    \end{tabular} 
  }
\end{table}

\begin{table}[htbp]
\centering{
  \topcaption{Systematic uncertainties for the subleading jet shape components, integrated over \xj, and $\Delta r$, and shown for \pp and centrality-binned PbPb collisions. The ranges correspond to the \pt dependence of the uncertainty. If some \pt bins have an uncertainty smaller than 0.5\%, the range is presented with a ``$<$" symbol and the upper bound.}
  \label{tab:subleadingJetSystematics}
  \begin{tabular}{cccccc}
    Source & 0--10\% &  10--30\% & 30--50\% & 50--90\% & \pp \\ 
    \hline
    Underlying event fluctuation bias & $<$1\% & $<$1\% & $<$1\% & $<$1\% & \NA \\
    Jet fragmentation bias  & $<$3\% & $<$2\% & $<$1\% & $<$1\% & $<$1\% \\ 
    Residual jet energy scale & 2--9\% & 2--9\% & 2--10\% & 2--10\% & 1--11\% \\
    Jet energy resolution & 1--3\% & 1--2\% & 1--2\% & 1--2\% & 1\% \\
    Tracking efficiency & 6\% & 6\% & 6\% & 6\% & 3\% \\
    Acceptance correction & $<$1\% & $<$1\% & $<$1\% & $<$1\% & $<$1\% \\
    Underlying event subtraction & $<$3\% & $<$3\% & $<$3\% & $<$3\% & $<$3\% \\[\cmsTabSkip]
    Total & 7--11\% & 7--11\% & 7--11\% & 7--11\% & 4--11\% \\
  \end{tabular}
}
\end{table}

\section{Results}
\label{sec:results}

\begin{figure}[p]
  \centering{
    \includegraphics[width=0.99\textwidth]{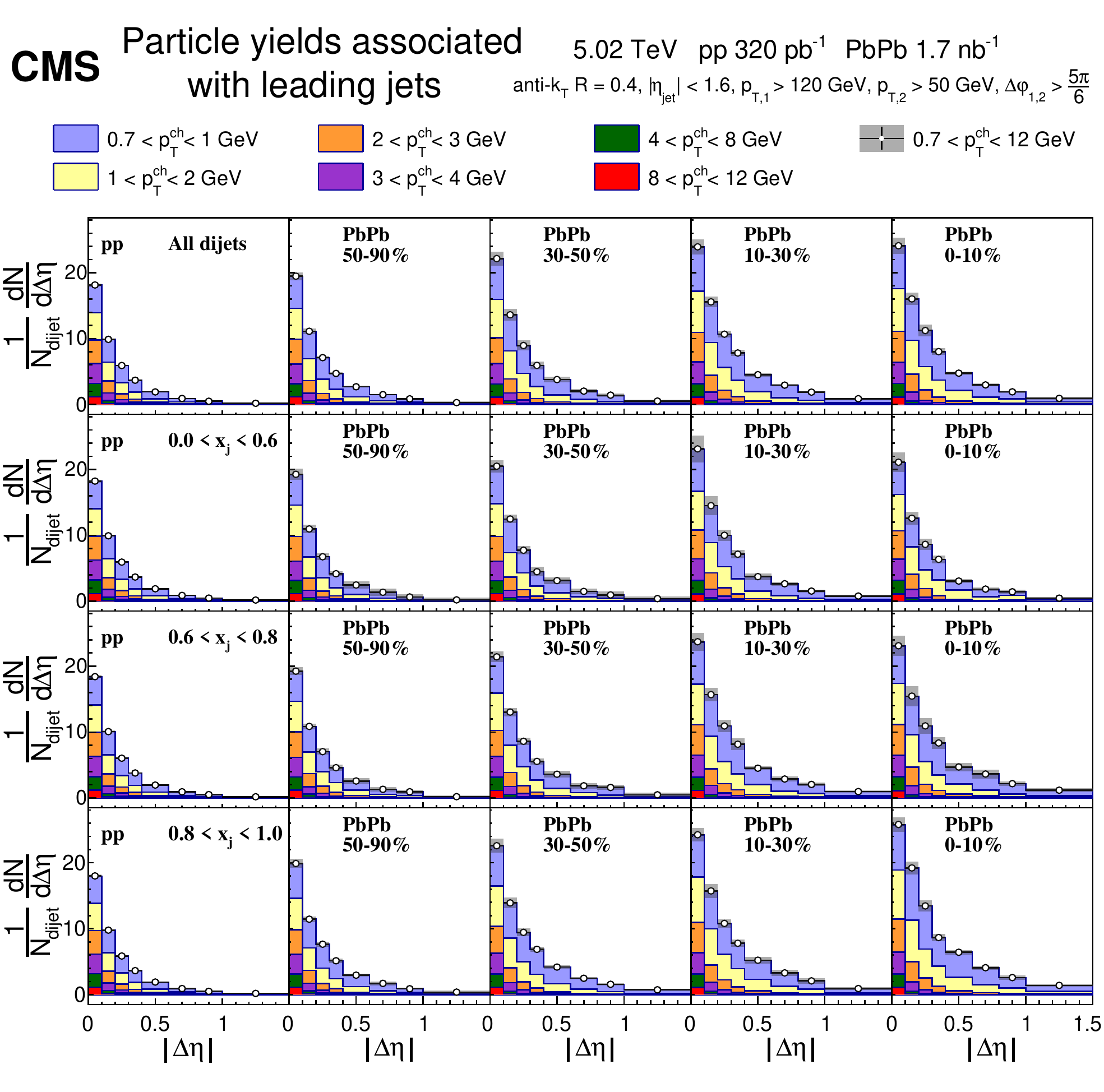}
    \caption{Distributions of charged-particle yields correlated to leading jets in the region $\abs{\Delta\varphi} < 1$ as a function of $\abs{\Delta\eta}$ for \pp (first column) and PbPb (second to fifth columns) collisions in different centrality bins, shown differentially for all \ptCh bins. The first row shows the charged-particle yields without any selection on \xj, while other rows show the charged-particle yields in different bins of \xj, starting with the most unbalanced $0 < \xj < 0.6$ (second row) to the most balanced $0.8 < \xj < 1.0$ (fourth row) dijet events.}
    \label{fig:yieldDeltaEtaLeading}
  }
\end{figure}

Figure~\ref{fig:yieldDeltaEtaLeading} shows the results for charged-particle yields in the region $\abs{\Delta\varphi} < 1$ as a function of $\abs{\Delta\eta}$ from the leading jets. The intervals in track \pt are indicated by the stacked histograms. The first row shows the charged-particle yields without any selection on \xj, while other rows show the charged-particle yields in different bins of \xj from the most unbalanced $0 < \xj < 0.6$ (second row) to the most balanced $0.8 < \xj < 1.0$ (fourth row) dijet events. The first panel in each row shows the charged-particle yields for the \pp collisions while other panels show the same yields for the PbPb collisions in different centrality bins, from the most peripheral 50--90\% (second panel) to the most central 0--10\% (fifth panel) collisions. Measurements of the per-jet invariant charged-particle yields show an enhancement of these yields in PbPb relative to \pp collisions. The enhancement is greatest for central collisions and decreases for more peripheral collisions. Comparing the different \xj bins, the enhancement of the total PbPb yield relative to \pp yield is seen to increase slightly as the dijet momenta become more balanced. A geometrical bias could explain this, as discussed, for example, in Ref.~\cite{Renk:jetProductionPoint}, since it can result in different path lengths inside the medium for the leading and subleading jets. In balanced dijet events, both jets lose significant amounts of energy, while in events with unbalanced dijet momenta, the leading jet is most likely produced closer to the surface of the plasma, thus losing less energy. The \xj trend in PbPb collisions could also be influenced by energy loss fluctuations, as studied in Ref.~\cite{asymmetryOrigins}.

\begin{figure}[hp!]
  \centering{
    \includegraphics[width=0.99\textwidth]{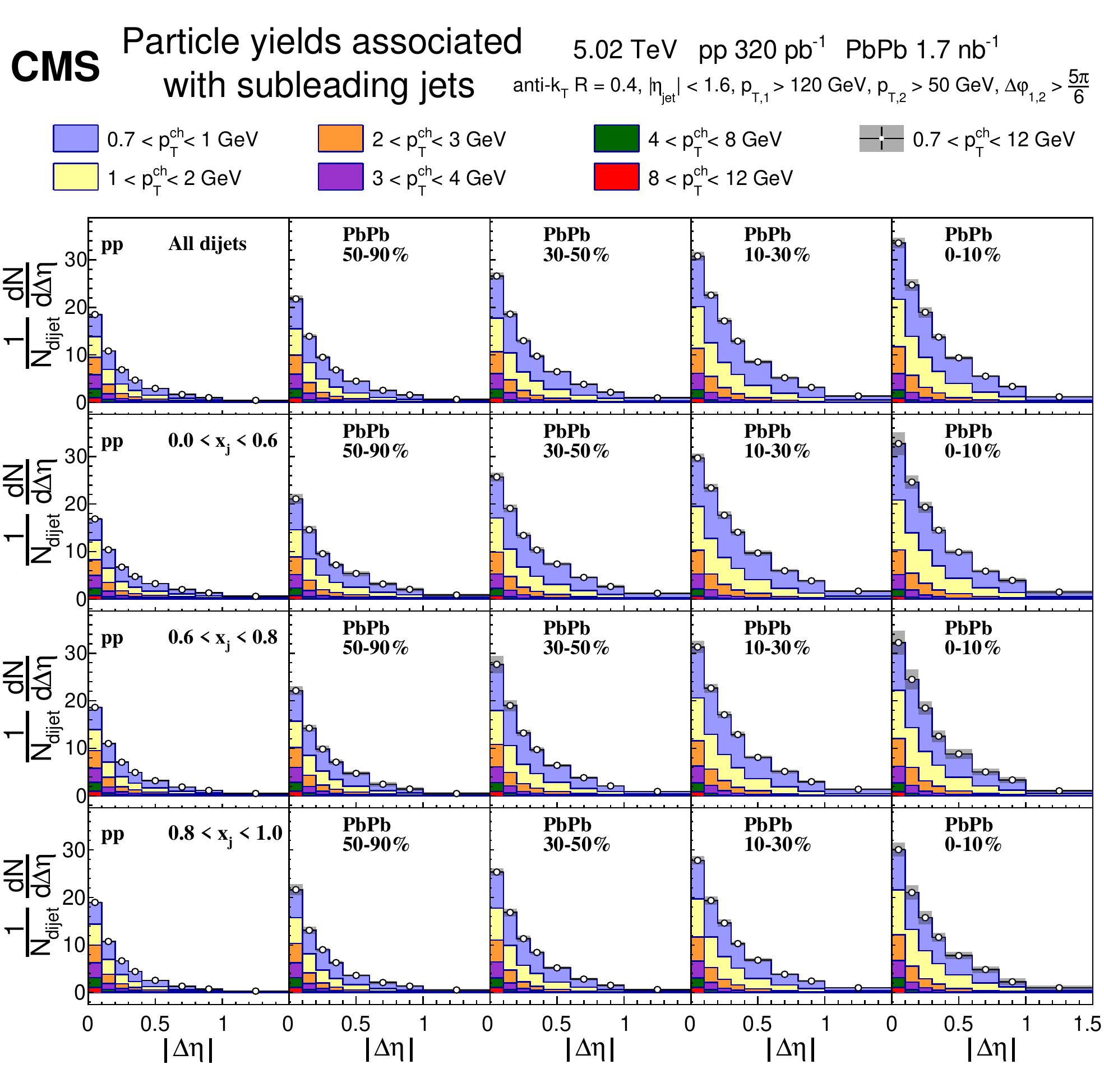}
    \caption{Distributions of charged-particle yields correlated to subleading jets in the region $\abs{\Delta\varphi} < 1$ as a function of $\abs{\Delta\eta}$ for \pp (first column) and PbPb (second to fifth columns) collisions in different centrality bins, shown differentially for all \ptCh bins. The first row shows the charged-particle yields without any selection on \xj, while other rows show the charged-particle yields in different bins of \xj, starting with the most unbalanced $0 < \xj < 0.6$ (second row) to the most balanced $0.8 < \xj < 1.0$ (fourth row) dijet events.}
    \label{fig:yieldDeltaEtaSubleading}
  }
\end{figure}

Figure~\ref{fig:yieldDeltaEtaSubleading} shows the results for charged-particle yields in the region $\abs{\Delta\varphi} < 1$ presented differentially in \ptCh, as a function of $\abs{\Delta\eta}$ for the subleading jets. The results are arranged in the same manner as for leading jets in Fig.~\ref{fig:yieldDeltaEtaLeading}. As with the leading jets, the measurements of the per-jet invariant charged-particle yields for subleading jets show an enhancement in PbPb relative to \pp collisions, with the greatest enhancement observed for central collisions. The magnitude of this enhancement is larger than for the leading jets in Fig.~\ref{fig:yieldDeltaEtaLeading} (note the different vertical scales between the two figures). Comparing the different \xj bins for the subleading jets, the enhancement of the total PbPb yield relative to the \pp yield is seen to slightly decrease as the dijet momenta become more balanced. This is opposite to the trend found for the leading jets and may reflect the greater path length through the plasma taken by the subleading jets for more unbalanced dijet events. When comparing leading and subleading results, there also might be an effect where the selection of leading and subleading jets in PbPb collisions is reversed as a consequence of parton energy loss in quark-gluon plasma.
 
The jet radial momentum distributions $\mathrm{P} (\Delta r)$ and jet shapes $\rho(\Delta r)$ are studied by examining the distribution of charged particles in annular rings of width $\delta r = 0.05$ around the jet axis. The jet radial momentum profile $\mathrm{P} (\Delta r)$ is the transverse momentum weighted distribution of particles around the jet axis: 
\begin{linenomath}
\begin{equation}
\mathrm{P} (\Delta r) = \frac{1}{\delta r} \frac{1}{\NText{jets}} \large{\Sigma_{\text{jets}}} \Sigma_{\text{tracks} \in (\Delta r_{a},\,\Delta r_{b})} \ptCh ,
\end{equation}
\end{linenomath}
where $\Delta r_{a}$ and $\Delta r_{b}$ define the annular edges of $\Delta r$, and $\delta r = \Delta r_{b} - \Delta r_{a}$. The jet shape $\rho(\Delta r)$ is the momentum distribution normalized to unity over $\Delta r < 1$, with
\begin{linenomath}
\begin{equation}
\rho(\Delta r) = \frac{\mathrm{P} (\Delta r)}{\Sigma_{\text{jets}} \Sigma_{\text{tracks} \in \Delta r < 1} \, \ptCh} .
\label{eq:jetShape}
\end{equation}
\end{linenomath}
The main difference between these two is that while jet shapes focus exclusively on studying the shape of the distribution, jet radial momentum distributions are sensitive to changes in the absolute scale of the in-jet momentum flow between \pp and PbPb collisions.

\begin{figure}[hp!]
\centering{
  \includegraphics[width=0.99\linewidth]{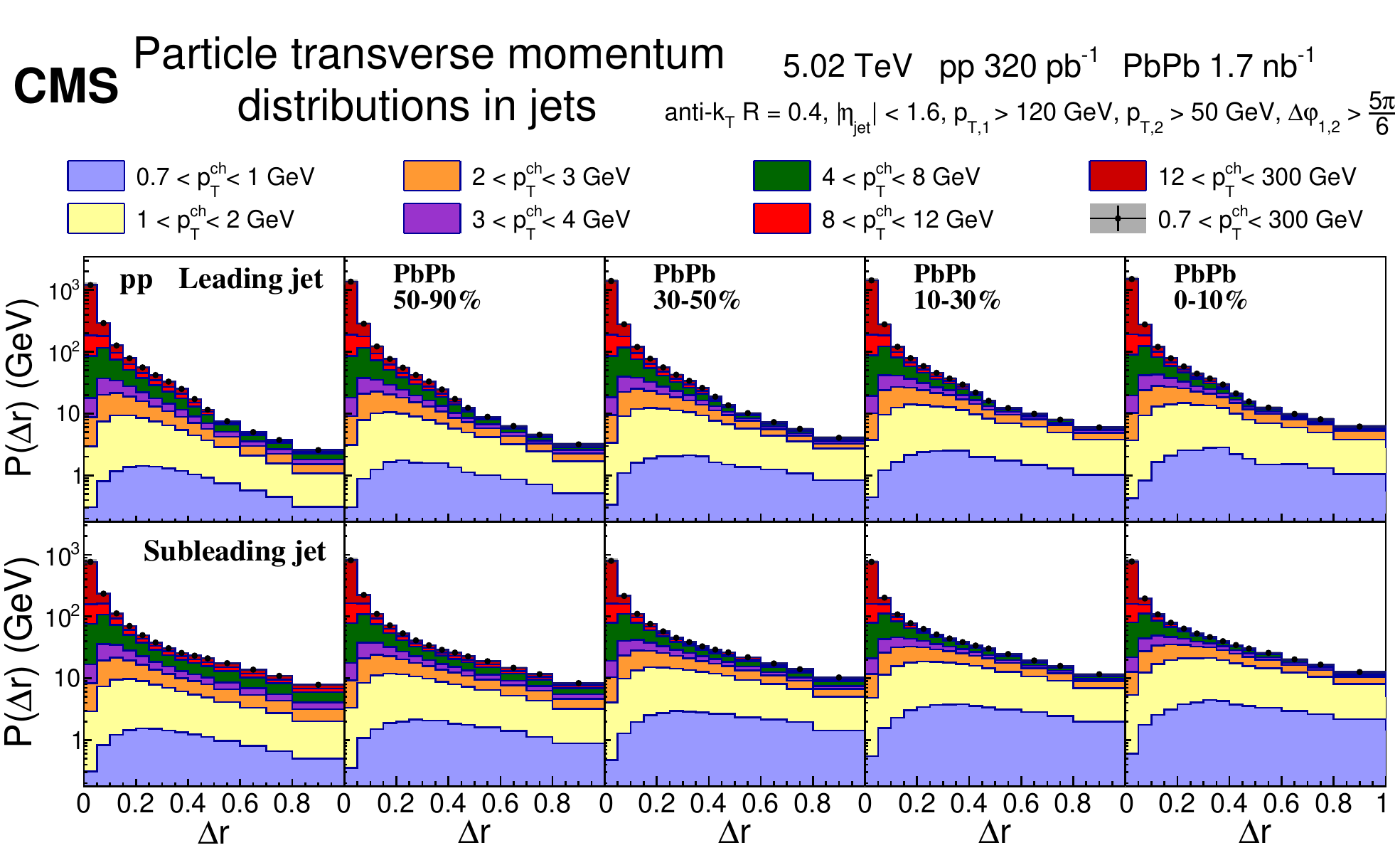}
\caption{Jet radial momentum profile $\mathrm{P}(\Delta r)$ for \pp (first column) and PbPb (second to fifth columns) collisions in different centrality bins as a function of $\Delta r$, shown differentially in \ptCh for leading (upper row) and subleading (lower row) jets.}
\label{fig:momentumProfileLeadingSubleadingJet}
}
\end{figure}

Figure~\ref{fig:momentumProfileLeadingSubleadingJet} shows the jet radial momentum distributions in PbPb and \pp collisions differentially in \ptCh. The first row shows the jet radial momentum distribution for the leading jets, while the second row is for subleading jets. The first panel in each row shows the results for \pp collisions, while other panels are for PbPb collisions in different centrality bins, starting from the most peripheral 50--90\% (second panel) to the most central 0--10\% (fifth panel) collisions. For both leading and subleading jets, when going toward more central events, the momentum profile at large $\Delta r$ is enhanced in PbPb collisions over the one in \pp collisions. The enhancement is largest for the low-\pt charged particles, as expected if the energy lost at high \pt resulting from interactions of partons with the quark-gluon plasma reappears in the form of low-\pt particles far away from the jet axis.

\begin{figure}[hp!]
\centering{
  \includegraphics[width=0.99\linewidth]{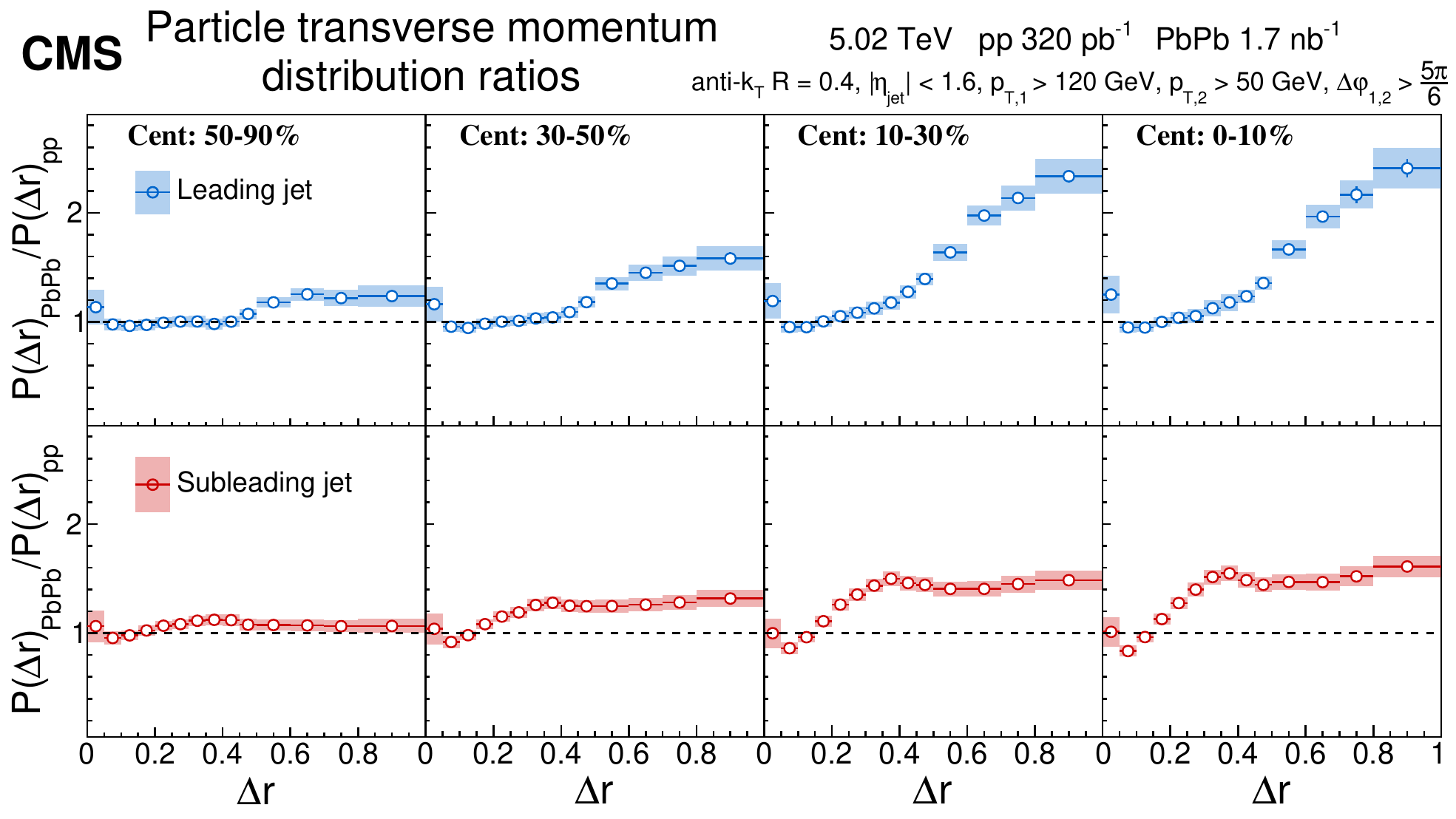}
\caption{The PbPb to \pp ratio of the jet radial momentum distributions as a function of $\Delta r$, $\mathrm{P} (\Delta r)_{\mathrm{PbPb}}/\mathrm{P} (\Delta r)_{\pp}$, for different centrality bins for the leading jets (upper row) and subleading jets (lower row).}
\label{fig:momentumProfileRatioLeadingSubleadingJet}
}
\end{figure}

Figure~\ref{fig:momentumProfileRatioLeadingSubleadingJet} shows the PbPb to \pp collision ratio of the jet radial momentum profiles for different centrality bins. A clear trend can be seen in these plots. The enhancement of the PbPb radial momentum distribution over the \pp distribution is the largest for the most central collisions and for larger separations of the charged particles from the jet axis. The enhancement for subleading jets is not as large as for the leading jets because of the widening of the \pp reference distribution. The main reason of this widening is the fact that subleading jets have significantly lower \pt compared to leading jets, and jet shapes for lower \pt jets are wider.

\begin{figure}[hp!]
  \centering{
    \includegraphics[width=0.99\linewidth]{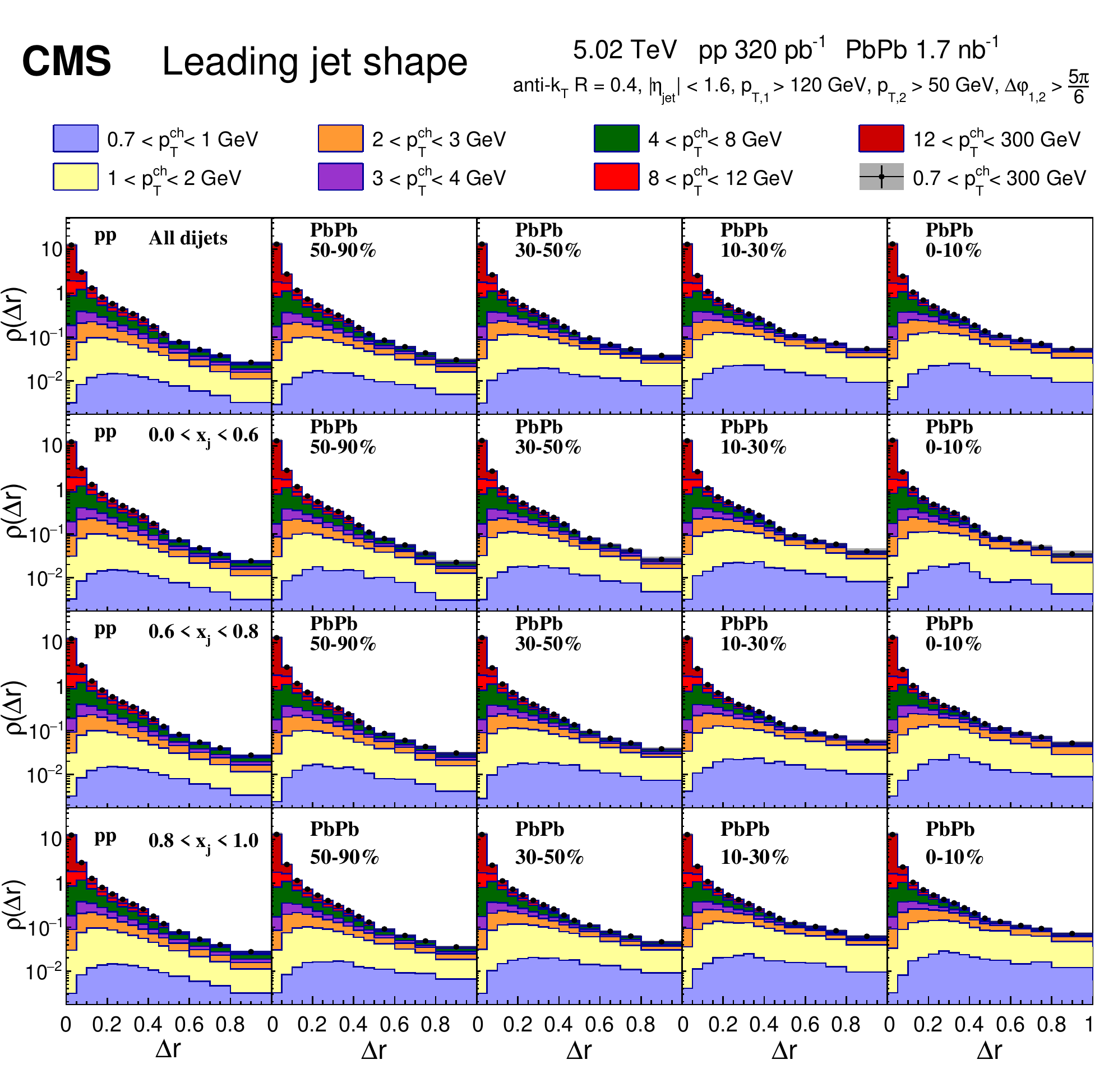}
    \caption{Leading jet shapes $\rho(\Delta r)$ (normalized to unity over $\Delta r < 1$) for \pp (first column) and PbPb (second to fifth columns) collisions in different centrality bins as a function of $\Delta r$, shown differentially in \ptCh for the inclusive \xj bin (first row) and in differential bins $0 < \xj < 0.6$ (second row),  $0.6 < \xj < 0.8$ (third row), and $0.8 < \xj < 1.0$ (fourth row).}
    \label{fig:jetShapeLeadingJet_Temp}
  }
\end{figure}

The jet shape results for the leading jets are presented in Fig.~\ref{fig:jetShapeLeadingJet_Temp}. The first row shows the jet shape without any selection on \xj, while other rows show the jet shape in different \xj bins. The first panel in each row shows the jet shape for \pp collisions. Without \xj selection, about 3.7\% of the total charged-particle constituent \pt is out-of-cone ($R > 0.4$). This does not change significantly as a function of \xj, varying between 3.5\% and 3.8\% from the most unbalanced to the most balanced studied \xj bins. The other panels are for PbPb collisions in different centrality bins. When compared to \pp collisions, there is an enhancement of low-\pt charged particles in PbPb collisions. This enhancement is greater for central events than for peripheral events.

\begin{figure}[hp!]
  \centering{
  \includegraphics[width=0.99\linewidth]{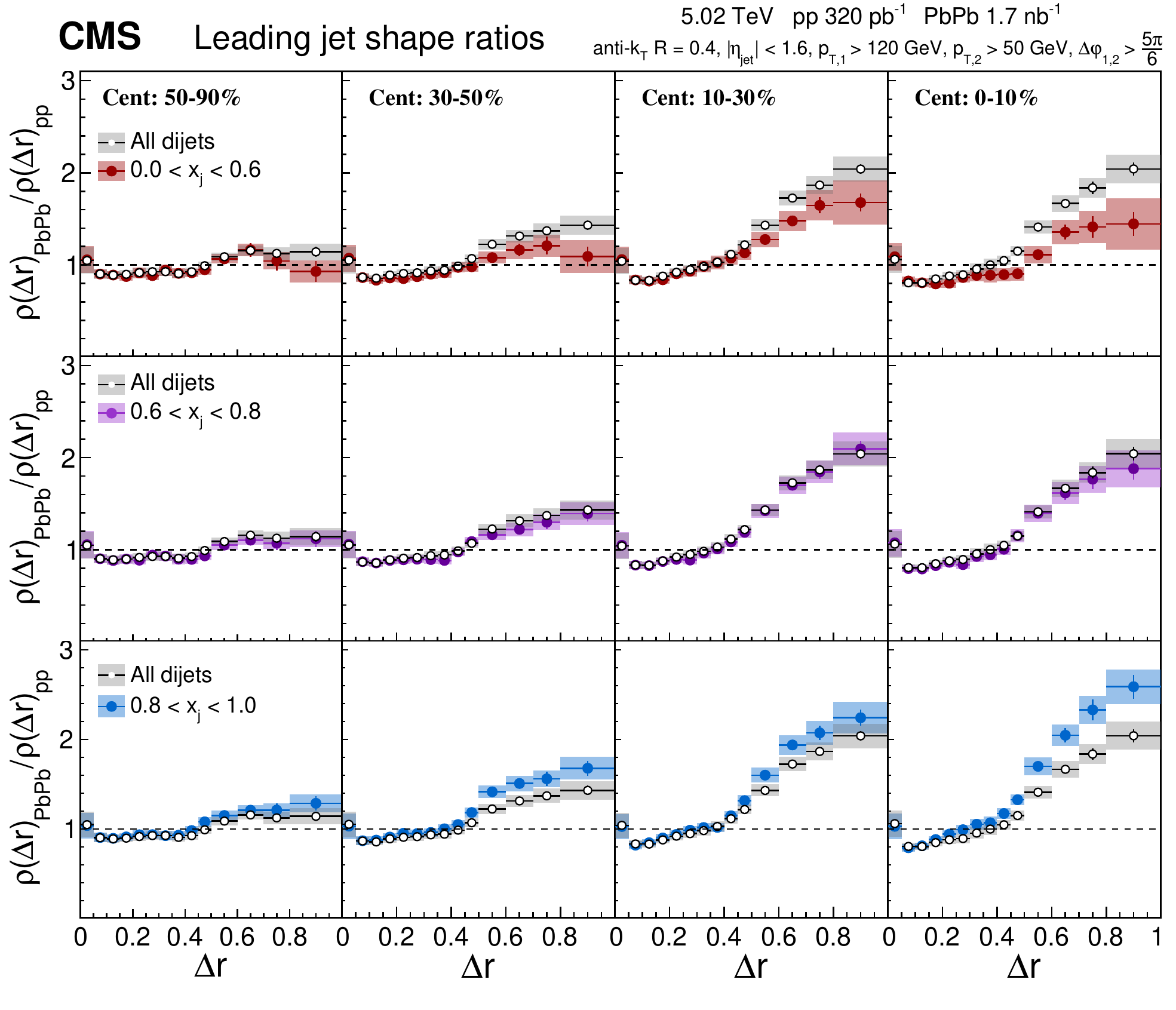}
  \caption{The PbPb to \pp ratio as a function of $\Delta r$ for leading jet shapes, $\rhoPbPb / \rhoPp$, in different centrality bins for $0 < \xj < 0.6$ (upper row), $0.6 < \xj < 0.8$ (middle row) and  $0.8 < \xj < 1.0$ (lower row) dijet selections. The leading jet shape ratio for all dijets, \ie, without any selection on the dijet momentum balance are also shown in each row for comparison. The error bars represent the statistical uncertainties and the shaded areas the systematic uncertainties.}
  \label{fig:jetShapeLeadingJet_ratio_Temp}
  }
\end{figure}

The modifications to the leading jet shapes in PbPb collisions are quantified in Fig.~\ref{fig:jetShapeLeadingJet_ratio_Temp}, which shows the $\rhoPbPb / \rhoPp$ ratios in different centrality and \xj bins. When going toward more central events from the peripheral ones, there is an enhancement of the PbPb jet shape compared to the \pp shape at large $\Delta r$. As already seen in the charged-particle yield plots of Fig.~\ref{fig:yieldDeltaEtaLeading}, the differences between \pp and PbPb results for leading jets are the largest for the most balanced collisions ($0.8 < \xj < 1.0$) and become smaller as the two jet momenta become less balanced. 

\begin{figure}[hp!]
\centering{
\includegraphics[width=0.99\linewidth]{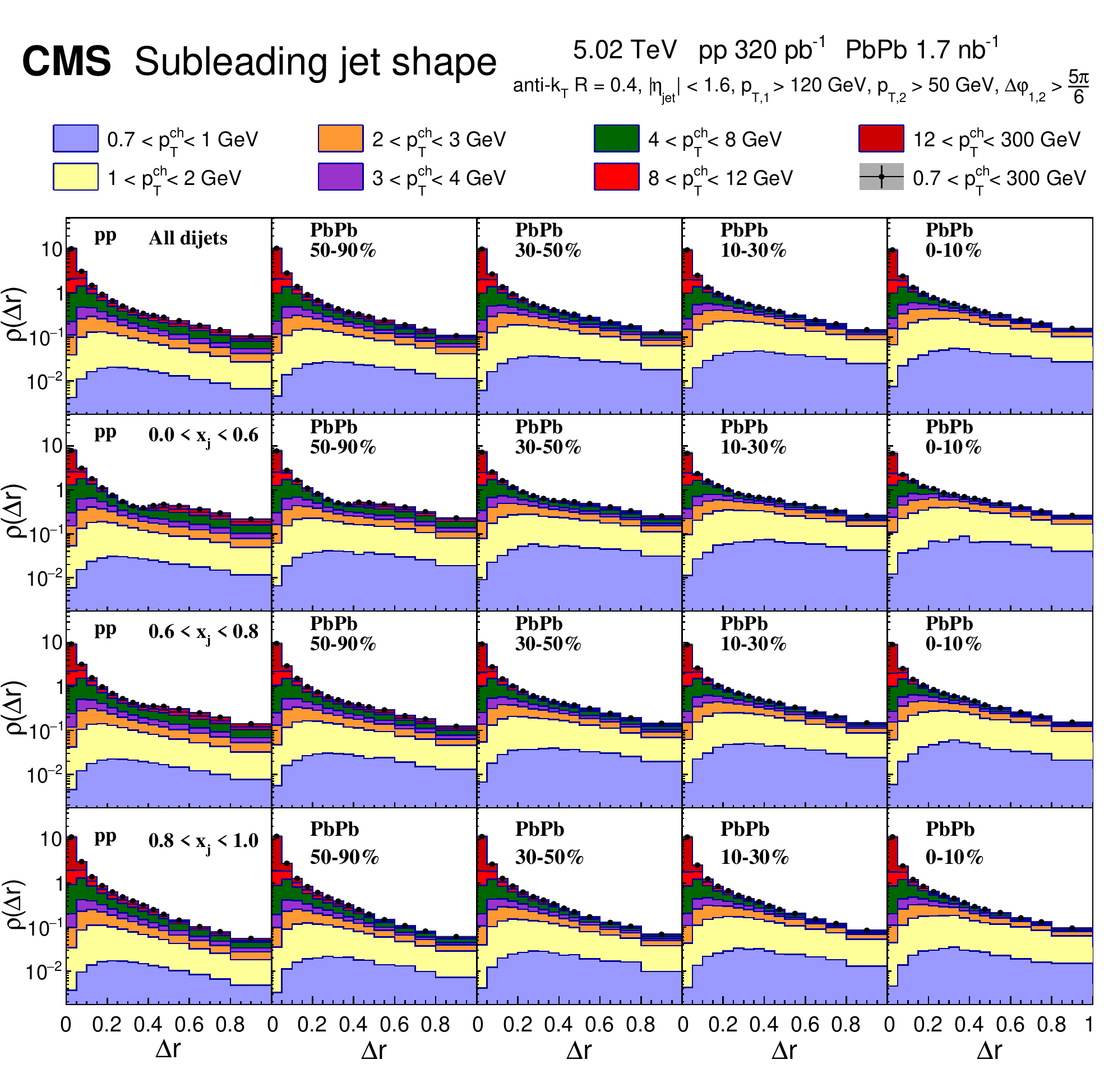}
\caption{Subleading jet shapes $\rho(\Delta r)$ (normalized to unity over $\Delta r < 1$) for \pp (first column) and PbPb (second to fifth columns) collisions in different centrality bins as a function of $\Delta r$, shown differentially in \ptCh for the inclusive \xj bin (first row) and in differential bins $0 < \xj < 0.6$ (second row),  $0.6 < \xj < 0.8$ (third row), and $0.8 < \xj < 1.0$ (fourth row).}
  \label{fig:jetShapeSubleadingJet_Temp}
}
\end{figure}

Figure~\ref{fig:jetShapeSubleadingJet_Temp} shows the jet shape results differentially in charged-particle \pt for the subleading jets, for different selections of centrality and \xj. As also found for the leading jet results, an enhancement of low-\pt charged particles in PbPb collisions is seen for the subleading jet. Also, the unbalanced \pp collision distribution is not monotonically decreasing toward large $\Delta r$, but there is an enhancement around $\Delta r \sim 0.5$, which is mostly caused by high-\pt particles. As the jet cone size used for this analysis is $R = 0.4$, these particles are not a part of the subleading jet. This enhancement is still somewhat visible in the peripheral PbPb collisions, but disappears for the central events. In the balanced collisions, no enhancement of high-\pt particles at large $\Delta r$ is visible. To create an unbalanced dijet configuration in \pp collisions, there is most often a third jet to conserve momentum, as confirmed by Monte Carlo studies and 3-jet selection in \pp events, which explains the enhancement of high-\pt particles outside of the jet cone. The observed features lead to a much more significant \xj dependence for the out-of-cone charged-particle constituent \pt fractions compared to those seen for leading jets. While only 7\% of the total charged-particle constituent \pt is out-of-cone for balanced dijets ($0.8 < \xj < 1$) in \pp collisions, this fraction increases to 20\% for unbalanced dijets ($0 < \xj < 0.6$). Without \xj selection, the fraction is about 11\%. In central heavy ion collisions, the jet energy can be lost to the medium and is seen as excess low-\pt particles at larger angles rather than as a third reconstructed jet.

\begin{figure}[hp!]
\centering{
  \includegraphics[width=0.99\linewidth]{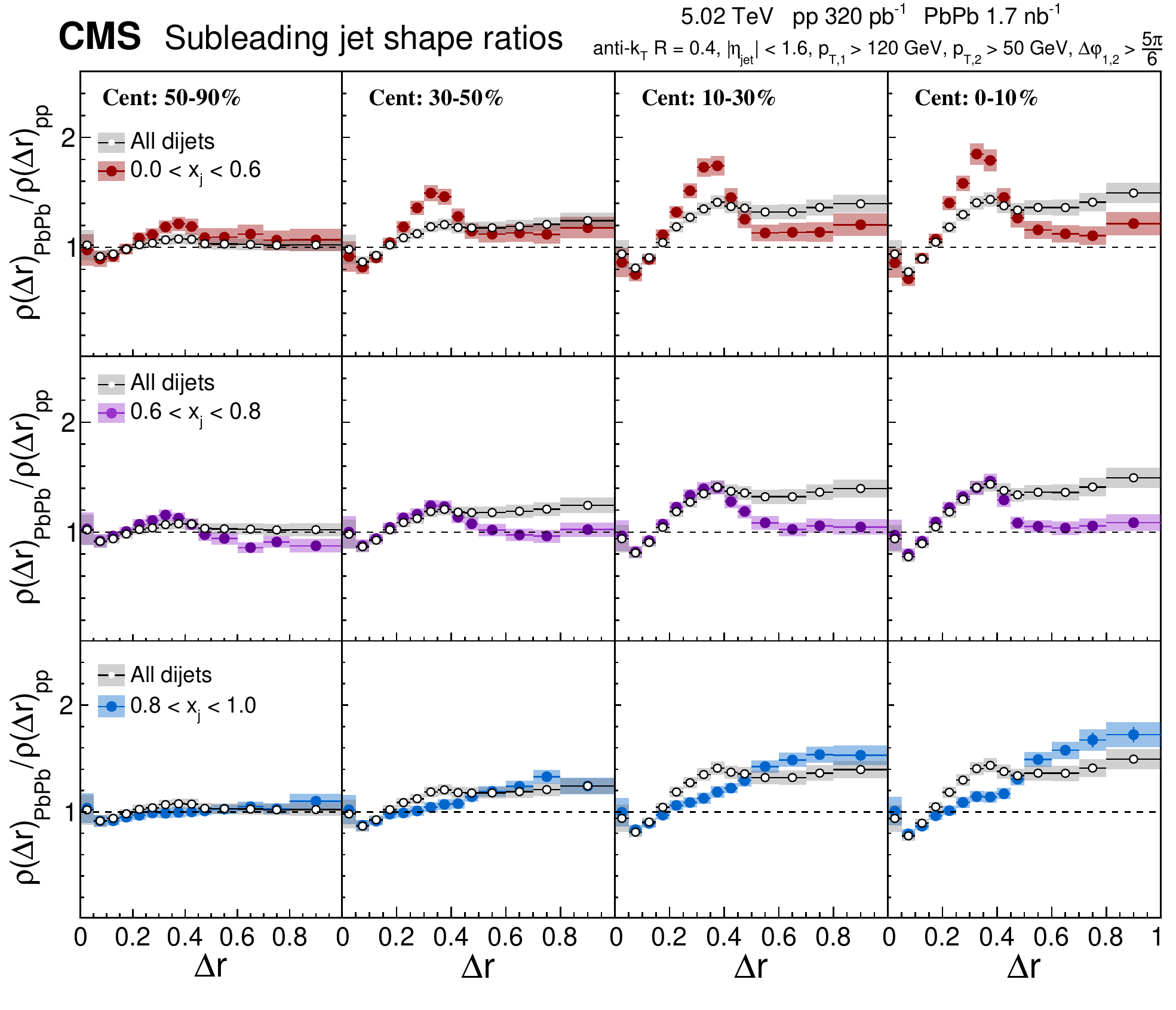}
  \caption{The PbPb to \pp ratio as a function of $\Delta r$ for subleading jet shapes, $\rhoPbPb / \rhoPp$, in different centrality bins for $0 < \xj < 0.6$ (upper row), $0.6 < \xj < 0.8$ (middle row) and  $0.8 < \xj < 1.0$ (lower row) dijet selections. The subleading jet shape ratio for all dijets, \ie, without any selection on the dijet momentum balance are also shown in each row for comparison. The error bars represent the statistical uncertainties and the shaded areas the systematic uncertainties.}   
  \label{fig:jetShapeSubleadingJet_ratio_Temp}
}
\end{figure}

The subleading jet shape ratios $\rhoPbPb / \rhoPp$ in different centrality and \xj bins are shown in Fig.~\ref{fig:jetShapeSubleadingJet_ratio_Temp}. There is a broadening of the PbPb jet shape compared to the \pp shape when going toward more central events from the peripheral ones. However, in the most unbalanced ($0.0 < \xj < 0.6$) and in moderately balanced ($0.6 < \xj < 0.8$) events, the ratio between PbPb and \pp jet shapes first peaks around $\Delta r = 0.4$, then gets close to one at large $\Delta r$. The enhancement from small $\Delta r$ toward the edge of the jet cone at $\Delta r = 0.4$ is explained by jet quenching. Outside the jet cone, the ratio gets closer to unity because of the reduced influence of any potential third jet in PbPb collisions, as discussed for Fig.~\ref{fig:jetShapeSubleadingJet_Temp}. If there would be no contribution from particles related to the third jet in the \pp jet shape, the enhancement would be the greatest at high $\Delta r$ also in these events, as is the case for the most balanced ($0.8 < \xj < 1.0$) events. Notice that the leading and subleading jet \pt can still be up to 20\% different in this most balanced bin. This explains the ratio differences in this bin between leading jets in Fig.~\ref{fig:jetShapeLeadingJet_ratio_Temp} and subleading jets in Fig.~\ref{fig:jetShapeSubleadingJet_ratio_Temp}.

\begin{figure}[hp!]
\centering{
  \includegraphics[width=0.99\linewidth]{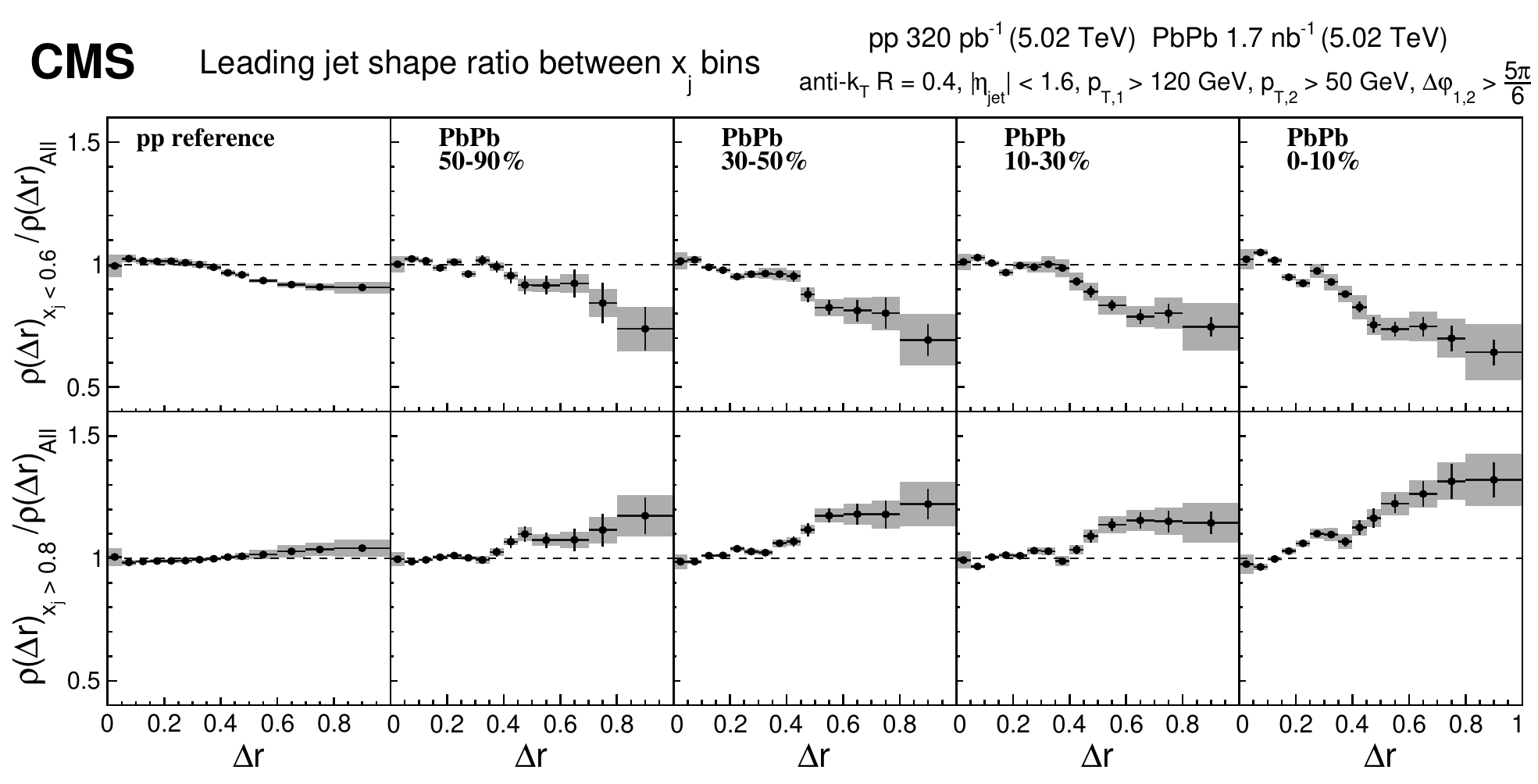}
\caption{Ratio of momentum-unbalanced ($0.0 < \xj < 0.6$, upper row) and balanced ($0.8 < \xj < 1.0$, lower row) jet shapes to \xj integrated jet shapes for leading jets in \pp collisions and different PbPb centrality bins as a function of $\Delta r$. The error bars represent the statistical uncertainties and the shaded areas the systematic uncertainties.}
\label{fig:jetShapeAsymmetryRatioLeading}
}
\end{figure}

\begin{figure}[hp!]
\centering{
  \includegraphics[width=0.99\linewidth]{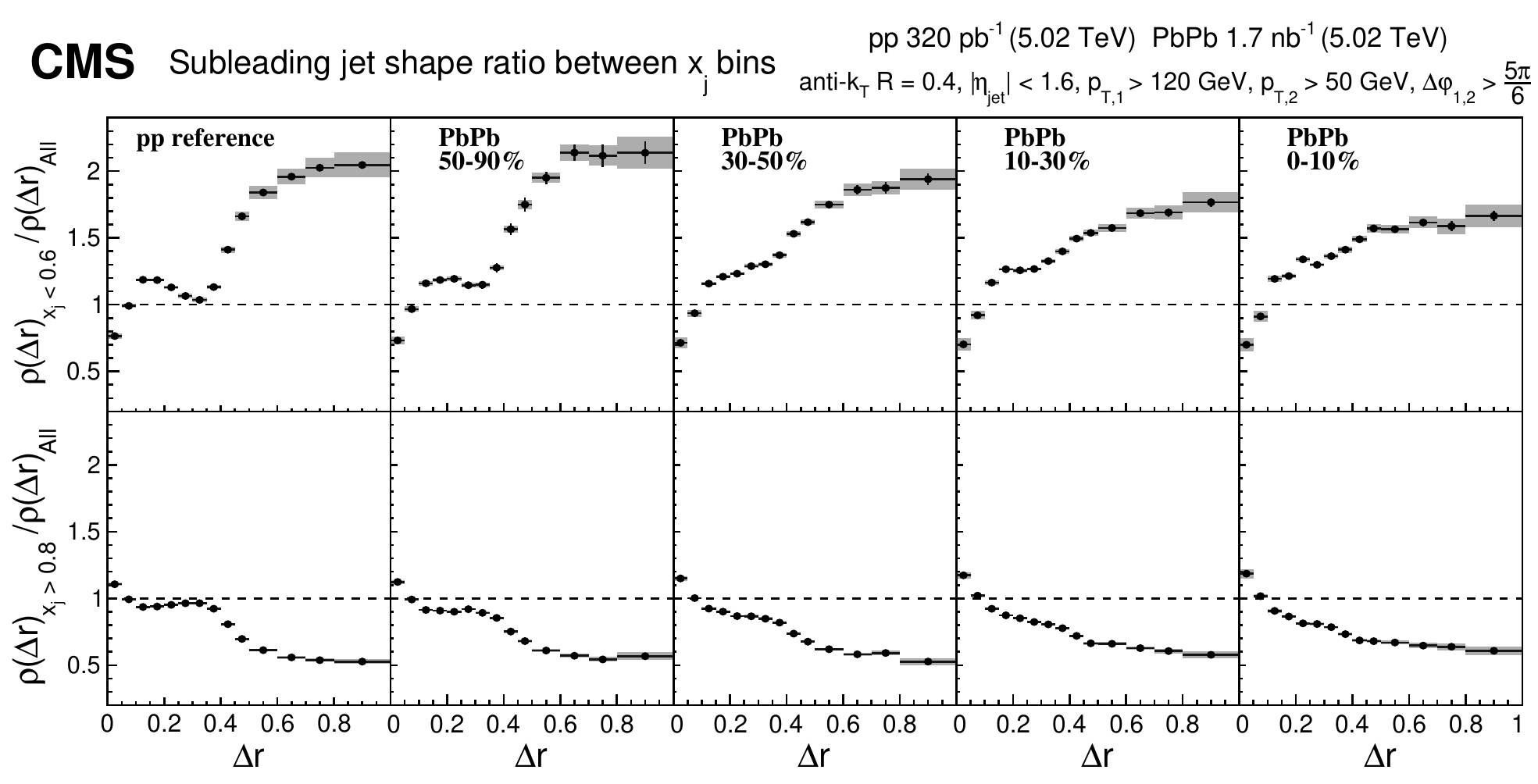}
\caption{Ratio of momentum-unbalanced ($0.0 < \xj < 0.6$, upper row) and balanced ($0.8 < \xj < 1.0$, lower row) jet shapes to \xj integrated jet shapes for subleading jets in \pp collisions and different PbPb centrality bins as a function of $\Delta r$. The error bars represent the statistical uncertainties and the shaded areas the systematic uncertainties.}
\label{fig:jetShapeAsymmetryRatioSubleading}
}
\end{figure}

The ratio of jet shapes in momentum balanced and unbalanced samples with the \xj inclusive sample for leading jets is presented in Fig.~\ref{fig:jetShapeAsymmetryRatioLeading} and for subleading jets in Fig.~\ref{fig:jetShapeAsymmetryRatioSubleading}. Taking this ratio cancels the systematic uncertainties related to tracking and the jet energy scale. However, the uncertainties related to the jet energy resolution do not cancel and are included in the systematic uncertainties. For the leading jets, the jet shapes in the unbalanced $0 < \xj < 0.6$ bin are narrower than in the \xj inclusive sample. Conversely, in the balanced $0.8 < \xj < 1.0$ bin, the jet shapes are wider. For the subleading jets in Fig.~\ref{fig:jetShapeAsymmetryRatioSubleading}, the behavior is exactly opposite to the leading jets, the jet shapes in the unbalanced bin are wider and in the balanced bin narrower compared to the \xj inclusive case. This is similar to the \xj dependence seen for the charged-particle distributions.

\section{Summary}
\label{sec:summary}

The CMS experiment has measured charged-particle yields and jet shapes in events containing back-to-back dijet pairs around the jet axes using data from proton-proton (\pp) and lead-lead (PbPb) collisions at $\sqrtsNN = 5.02\TeV$ collected in 2017 and 2018. When comparing the charged-particle yields around the jet axes, an excess of low transverse momentum (\pt) particles is observed in PbPb with respect to \pp collisions. This excess is larger for subleading jets compared to leading jets. The excess is also found to have a different dijet momentum balance $\xj = \ptSub / \ptLead$ dependence for the leading and subleading jets. A possible cause for \xj imbalance is that the leading jet is produced near the surface of the quark-gluon plasma while the subleading jet needs to traverse a longer distance through the plasma. The leading jets show the strongest modifications in balanced events ($\xj \approx 1$), while subleading jets experience the most pronounced modifications in events with a large jet momentum imbalance. A possible explanation is that in balanced events both jets lose a comparable amount of energy, while in events with a momentum imbalance the leading jet loses significantly less energy. Furthermore, jet quenching could lead to the reversal of the apparent leading/subleading ordering and energy loss fluctuations in PbPb collisions could also play a role.

For the jet shapes, which are normalized distributions of charged-particle transverse momentum as a function of the distance from the jet axis, a redistribution of energy is observed from small angles with respect to the jet axis to larger angles when comparing PbPb and \pp events. The difference between the PbPb and \pp results is larger for the leading jets compared to the subleading jets, which can be explained by the subleading jet distribution in \pp collisions being significantly wider than that for leading jets. When studying the unbalanced \xj selection for the subleading jets in \pp collisions, a fragmentation pattern consistent with the presence of a third jet is observed. Such a pattern is not observed in balanced \pp events or in the PbPb sample. As a result, the enhancement of the PbPb to \pp ratio for unbalanced events peaks around $\Delta r = 0.4$ and becomes smaller at larger $\Delta r$.

When comparing the jet shapes corresponding to different dijet momentum balance conditions, the distributions for leading jets are found to be the widest for events with balanced jet momenta. For subleading jets, the situation is the opposite, and the widest distributions are found from events having a significant momentum imbalance. These observations are consistent with the interpretation of the charged-particle yield measurements, namely that the average path length inside the medium for leading jets is larger for momentum balanced events, while for subleading jets it is larger in unbalanced events. By studying the charged-particle yields correlated to jets and jet shapes for the first time as a function of dijet momentum balance, this study provides new constraints to the theoretical models
and provides a unique way to explore the transition between the domains of weakly and strongly interacting QCD matter.

\begin{acknowledgments}
  We congratulate our colleagues in the CERN accelerator departments for the excellent performance of the LHC and thank the technical and administrative staffs at CERN and at other CMS institutes for their contributions to the success of the CMS effort. In addition, we gratefully acknowledge the computing centres and personnel of the Worldwide LHC Computing Grid and other centres for delivering so effectively the computing infrastructure essential to our analyses. Finally, we acknowledge the enduring support for the construction and operation of the LHC, the CMS detector, and the supporting computing infrastructure provided by the following funding agencies: BMBWF and FWF (Austria); FNRS and FWO (Belgium); CNPq, CAPES, FAPERJ, FAPERGS, and FAPESP (Brazil); MES (Bulgaria); CERN; CAS, MoST, and NSFC (China); COLCIENCIAS (Colombia); MSES and CSF (Croatia); RIF (Cyprus); SENESCYT (Ecuador); MoER, ERC PUT and ERDF (Estonia); Academy of Finland, MEC, and HIP (Finland); CEA and CNRS/IN2P3 (France); BMBF, DFG, and HGF (Germany); GSRT (Greece); NKFIA (Hungary); DAE and DST (India); IPM (Iran); SFI (Ireland); INFN (Italy); MSIP and NRF (Republic of Korea); MES (Latvia); LAS (Lithuania); MOE and UM (Malaysia); BUAP, CINVESTAV, CONACYT, LNS, SEP, and UASLP-FAI (Mexico); MOS (Montenegro); MBIE (New Zealand); PAEC (Pakistan); MSHE and NSC (Poland); FCT (Portugal); JINR (Dubna); MON, RosAtom, RAS, RFBR, and NRC KI (Russia); MESTD (Serbia); SEIDI, CPAN, PCTI, and FEDER (Spain); MOSTR (Sri Lanka); Swiss Funding Agencies (Switzerland); MST (Taipei); ThEPCenter, IPST, STAR, and NSTDA (Thailand); TUBITAK and TAEK (Turkey); NASU (Ukraine); STFC (United Kingdom); DOE and NSF (USA).
   
  \hyphenation{Rachada-pisek} Individuals have received support from the Marie-Curie programme and the European Research Council and Horizon 2020 Grant, contract Nos.\ 675440, 724704, 752730, and 765710 (European Union); the Leventis Foundation; the Alfred P.\ Sloan Foundation; the Alexander von Humboldt Foundation; the Belgian Federal Science Policy Office; the Fonds pour la Formation \`a la Recherche dans l'Industrie et dans l'Agriculture (FRIA-Belgium); the Agentschap voor Innovatie door Wetenschap en Technologie (IWT-Belgium); the F.R.S.-FNRS and FWO (Belgium) under the ``Excellence of Science -- EOS" -- be.h project n.\ 30820817; the Beijing Municipal Science \& Technology Commission, No. Z191100007219010; the Ministry of Education, Youth and Sports (MEYS) of the Czech Republic; the Deutsche Forschungsgemeinschaft (DFG), under Germany's Excellence Strategy -- EXC 2121 ``Quantum Universe" -- 390833306, and under project number 400140256 - GRK2497; the Lend\"ulet (``Momentum") Programme and the J\'anos Bolyai Research Scholarship of the Hungarian Academy of Sciences, the New National Excellence Program \'UNKP, the NKFIA research grants 123842, 123959, 124845, 124850, 125105, 128713, 128786, and 129058 (Hungary); the Council of Science and Industrial Research, India; the HOMING PLUS programme of the Foundation for Polish Science, cofinanced from European Union, Regional Development Fund, the Mobility Plus programme of the Ministry of Science and Higher Education, the National Science Center (Poland), contracts Harmonia 2014/14/M/ST2/00428, Opus 2014/13/B/ST2/02543, 2014/15/B/ST2/03998, and 2015/19/B/ST2/02861, Sonata-bis 2012/07/E/ST2/01406; the National Priorities Research Program by Qatar National Research Fund; the Ministry of Science and Higher Education, project no. 0723-2020-0041 (Russia); the Tomsk Polytechnic University Competitiveness Enhancement Program; the Programa Estatal de Fomento de la Investigaci{\'o}n Cient{\'i}fica y T{\'e}cnica de Excelencia Mar\'{\i}a de Maeztu, grant MDM-2015-0509 and the Programa Severo Ochoa del Principado de Asturias; the Thalis and Aristeia programmes cofinanced by EU-ESF and the Greek NSRF; the Rachadapisek Sompot Fund for Postdoctoral Fellowship, Chulalongkorn University and the Chulalongkorn Academic into Its 2nd Century Project Advancement Project (Thailand); the Kavli Foundation; the Nvidia Corporation; the SuperMicro Corporation; the Welch Foundation, contract C-1845; and the Weston Havens Foundation (USA).\end{acknowledgments}

\bibliography{auto_generated}
\clearpage
\appendix
\numberwithin{figure}{section}
\section{Dijet momentum balance migration matrices}
\label{app:Matrices}

\begin{figure}[h]
  \centering{
    \includegraphics[width=0.32\linewidth]{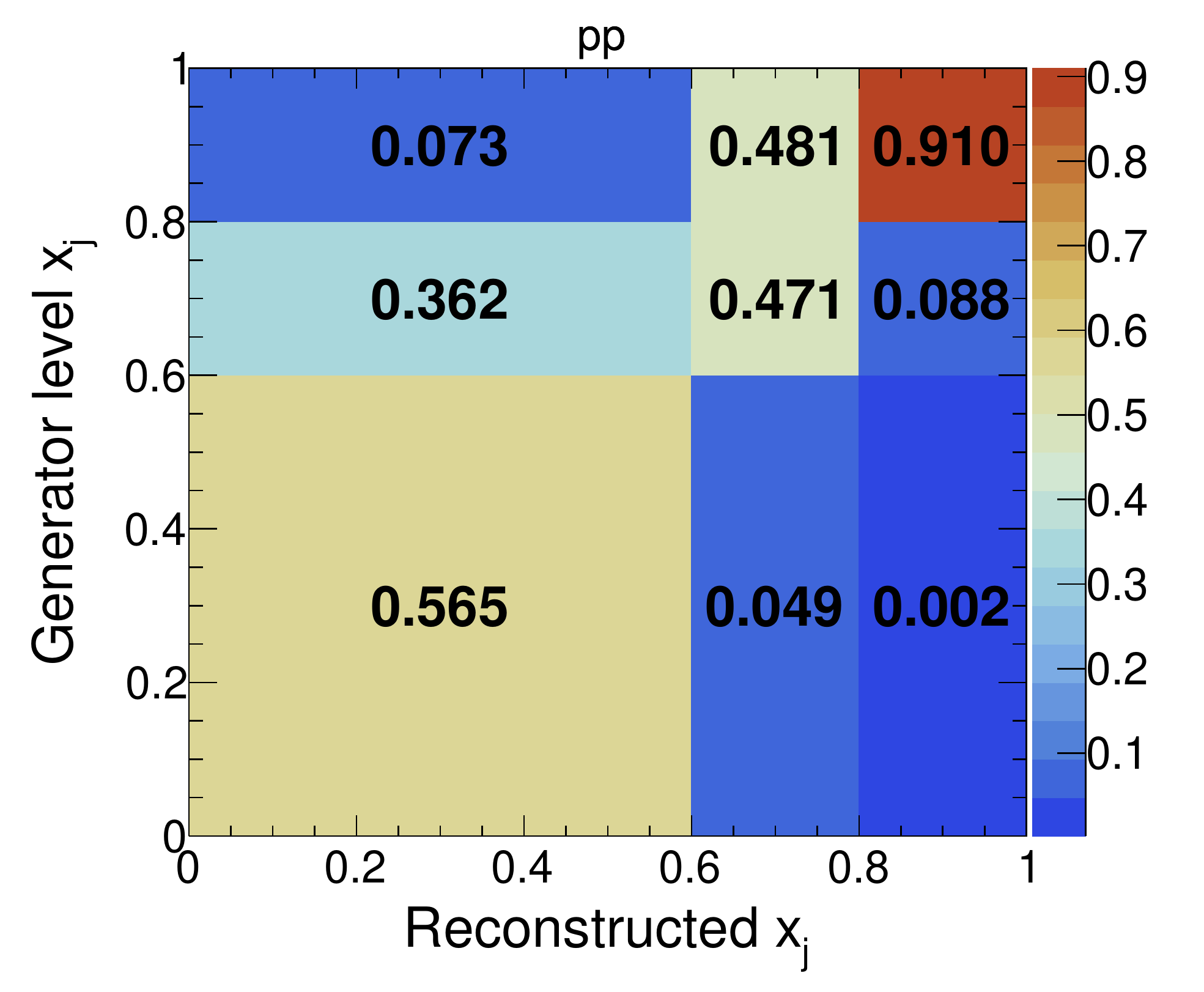} 
    \includegraphics[width=0.32\linewidth]{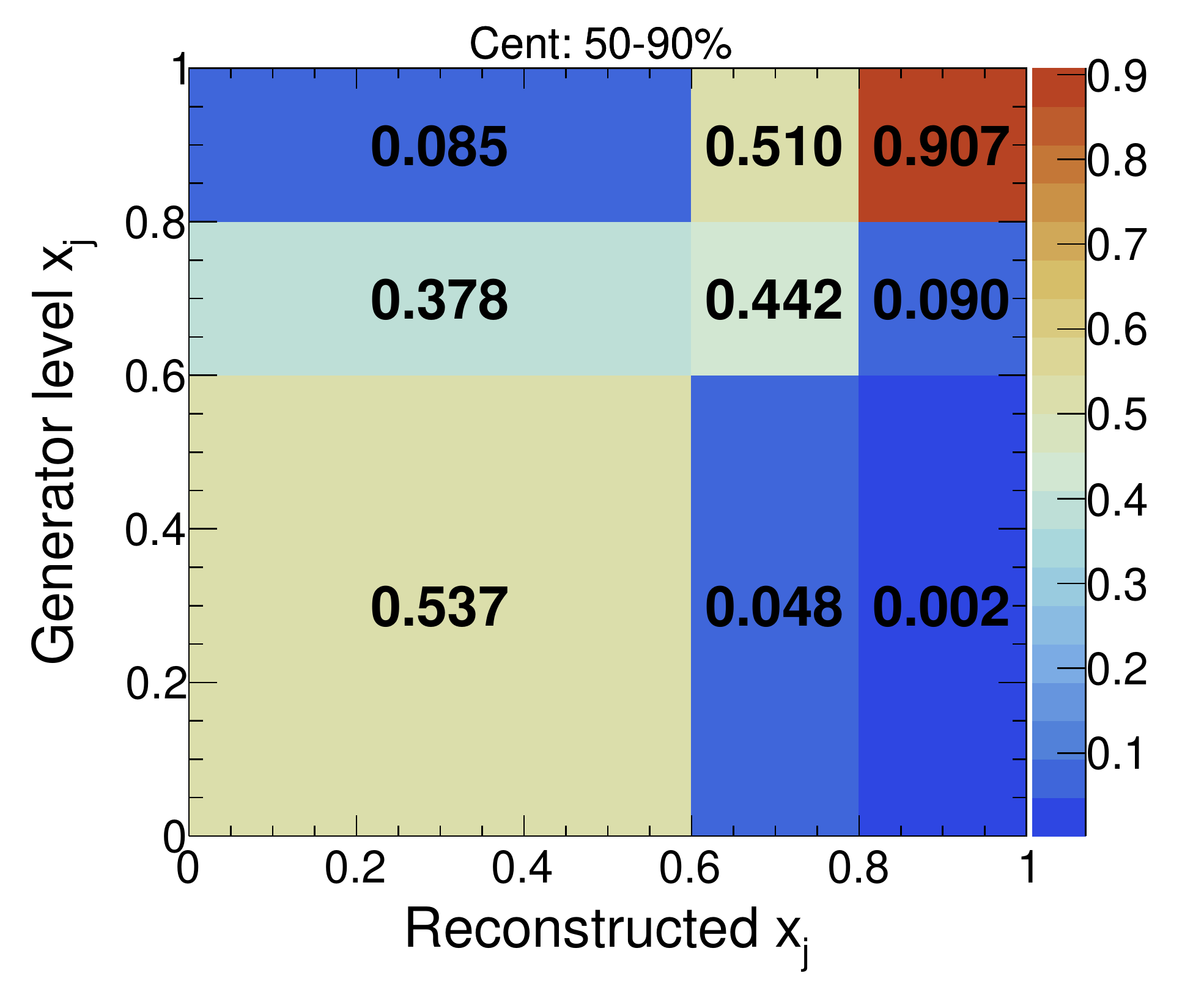}
    \includegraphics[width=0.32\linewidth]{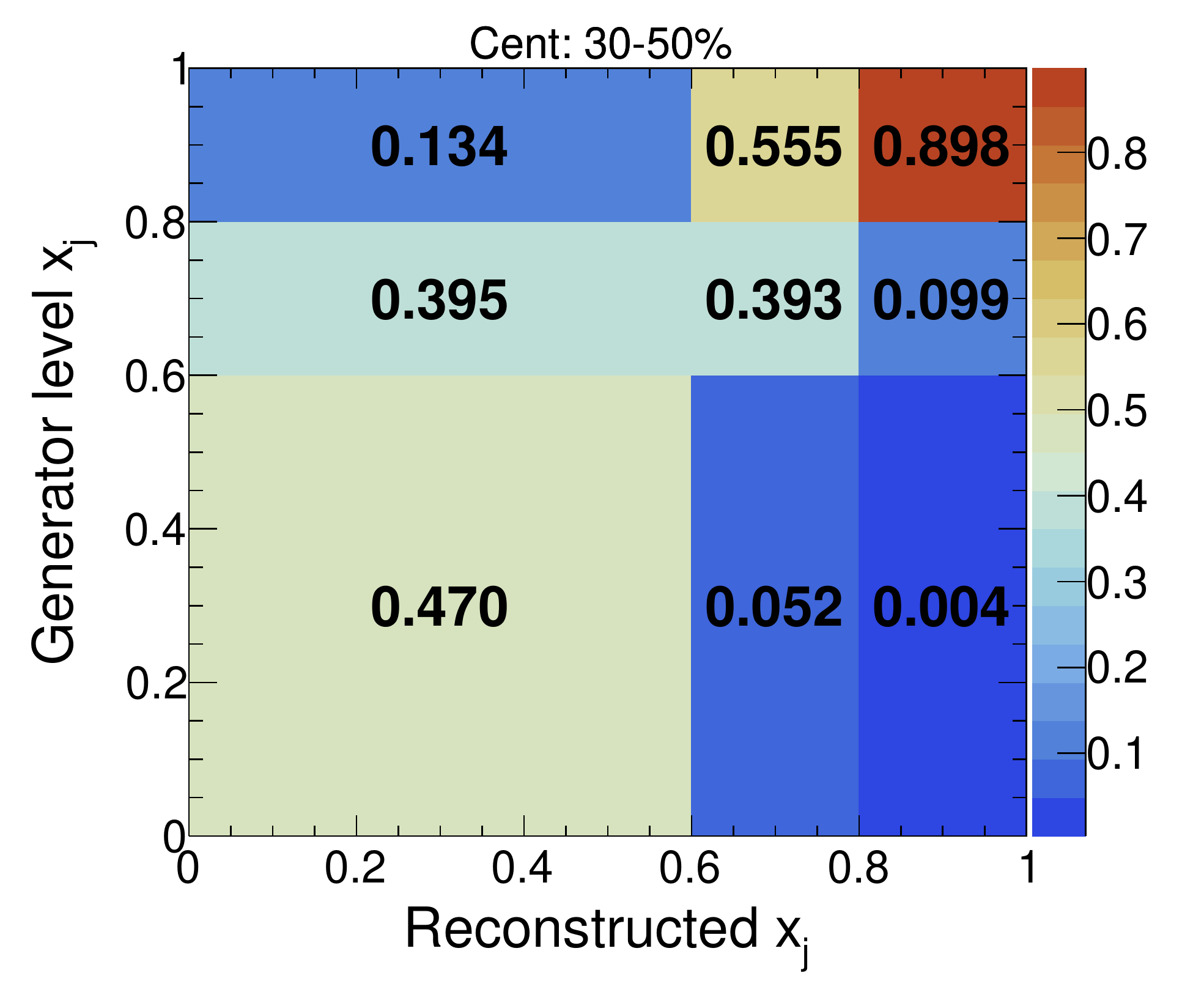} \\
    \includegraphics[width=0.32\linewidth]{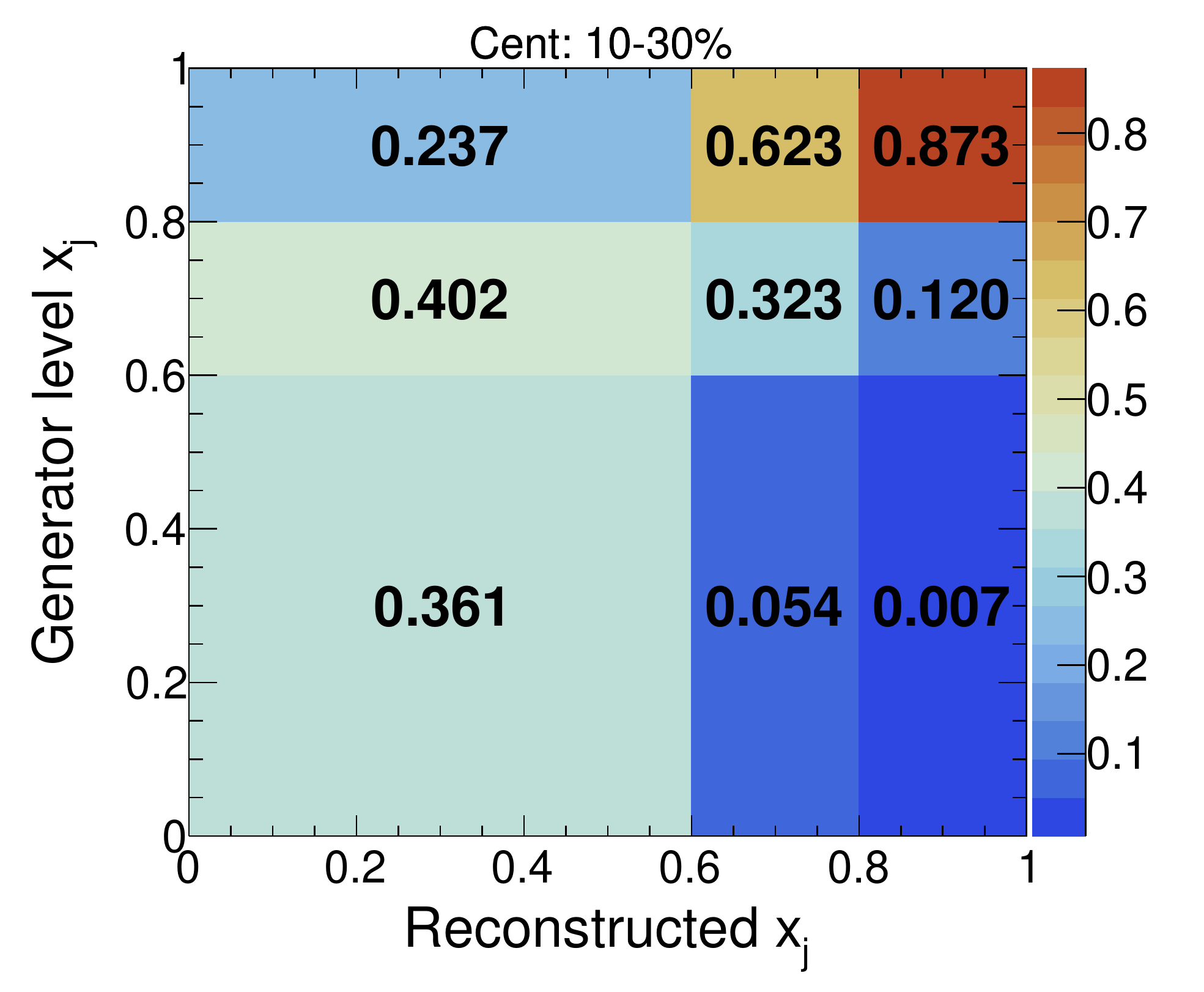}
    \includegraphics[width=0.32\linewidth]{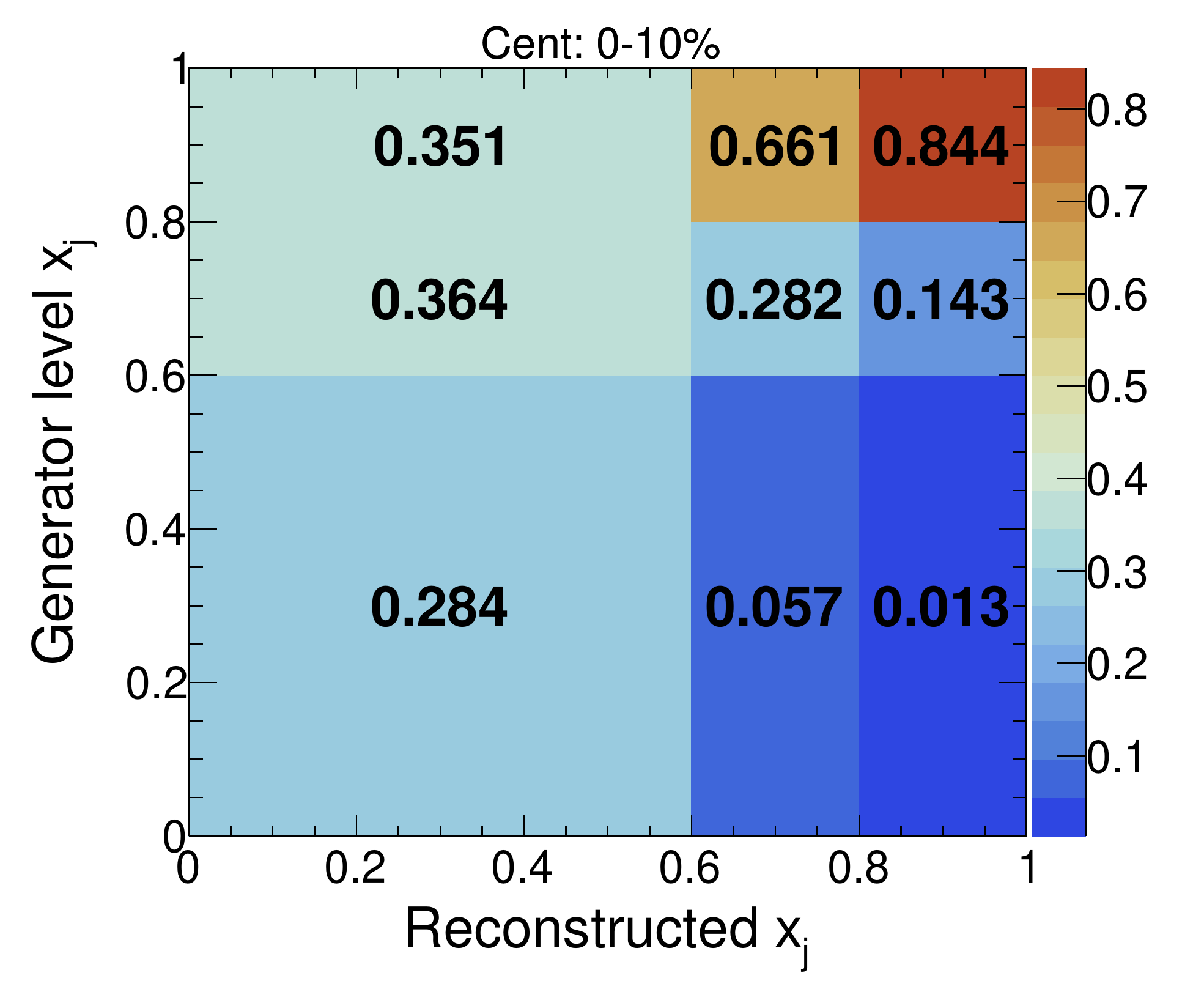} 
    \caption{Generator-level vs.\ reconstructed \xj values in the analysis \xj bins. The plots show the probability to find a generator level \xj for a given reconstructed \xj. The \PYTHIA~8 simulation is shown in the upper-left plot while the most central {\PYTHIA}+\HYDJET is shown in the lower-right plot.}
  \label{fig:xjMatrixWide}
  }
\end{figure}

\begin{figure}[h]
  \centering{
    \includegraphics[width=0.32\linewidth]{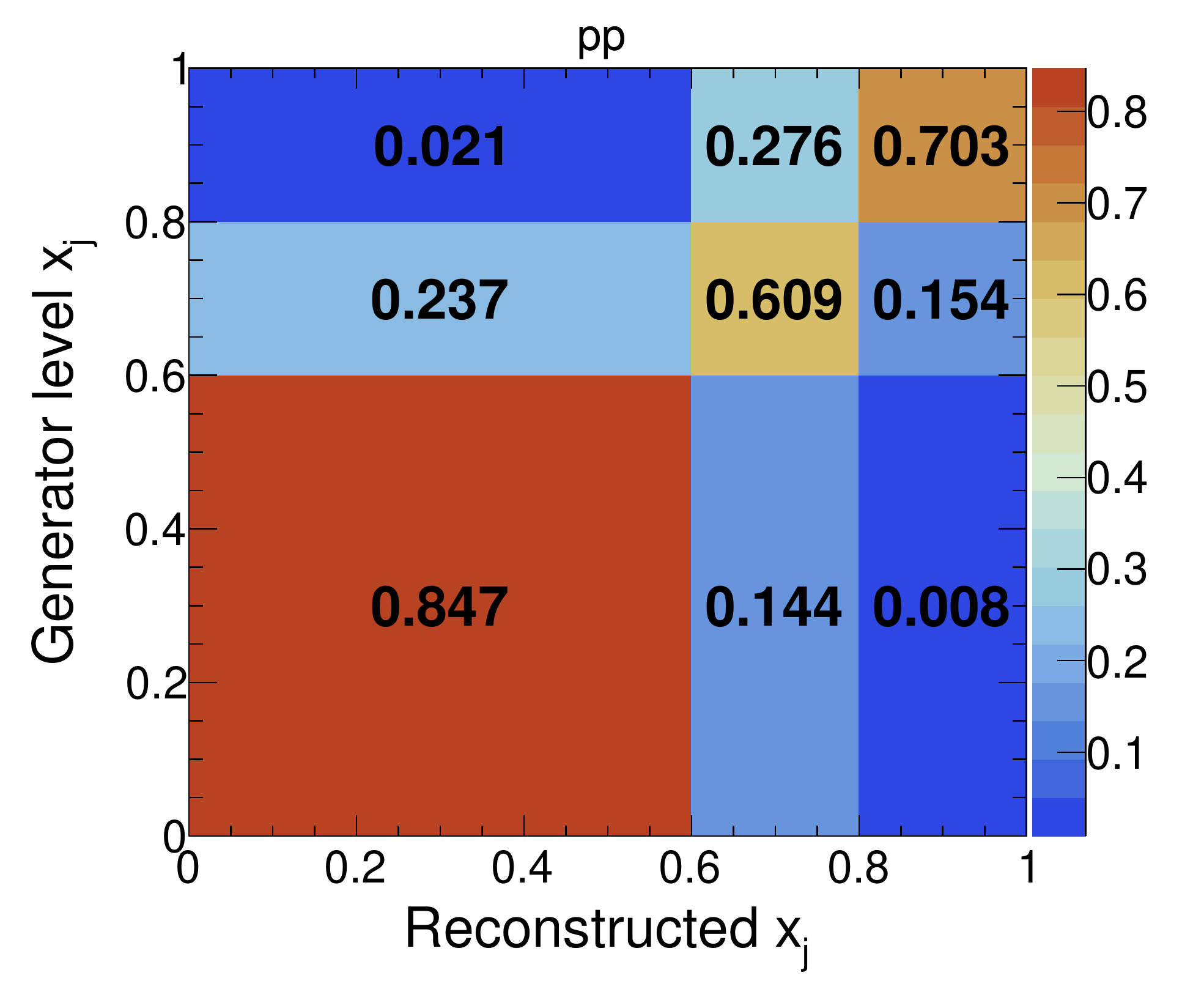} 
    \includegraphics[width=0.32\linewidth]{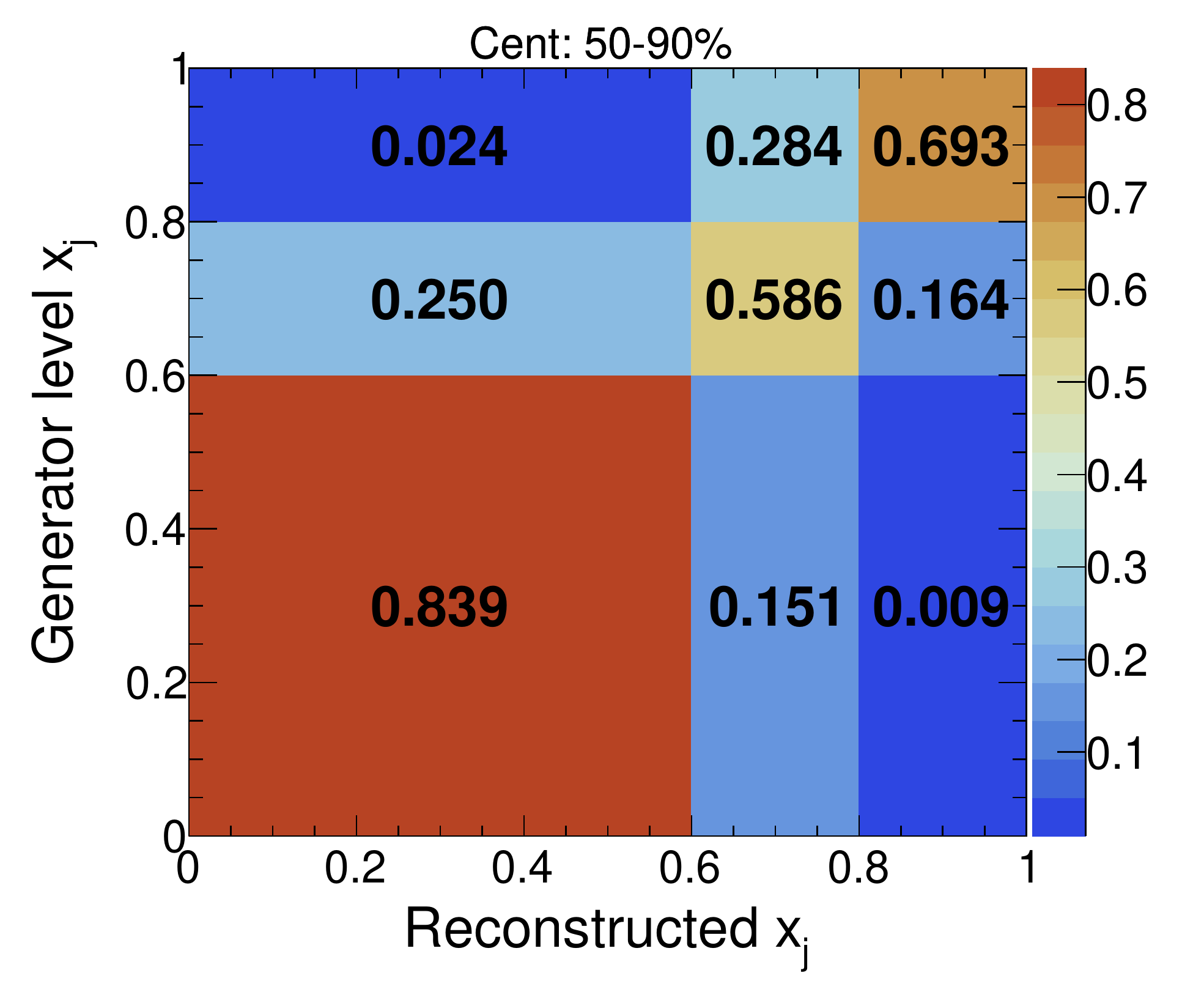}
    \includegraphics[width=0.32\linewidth]{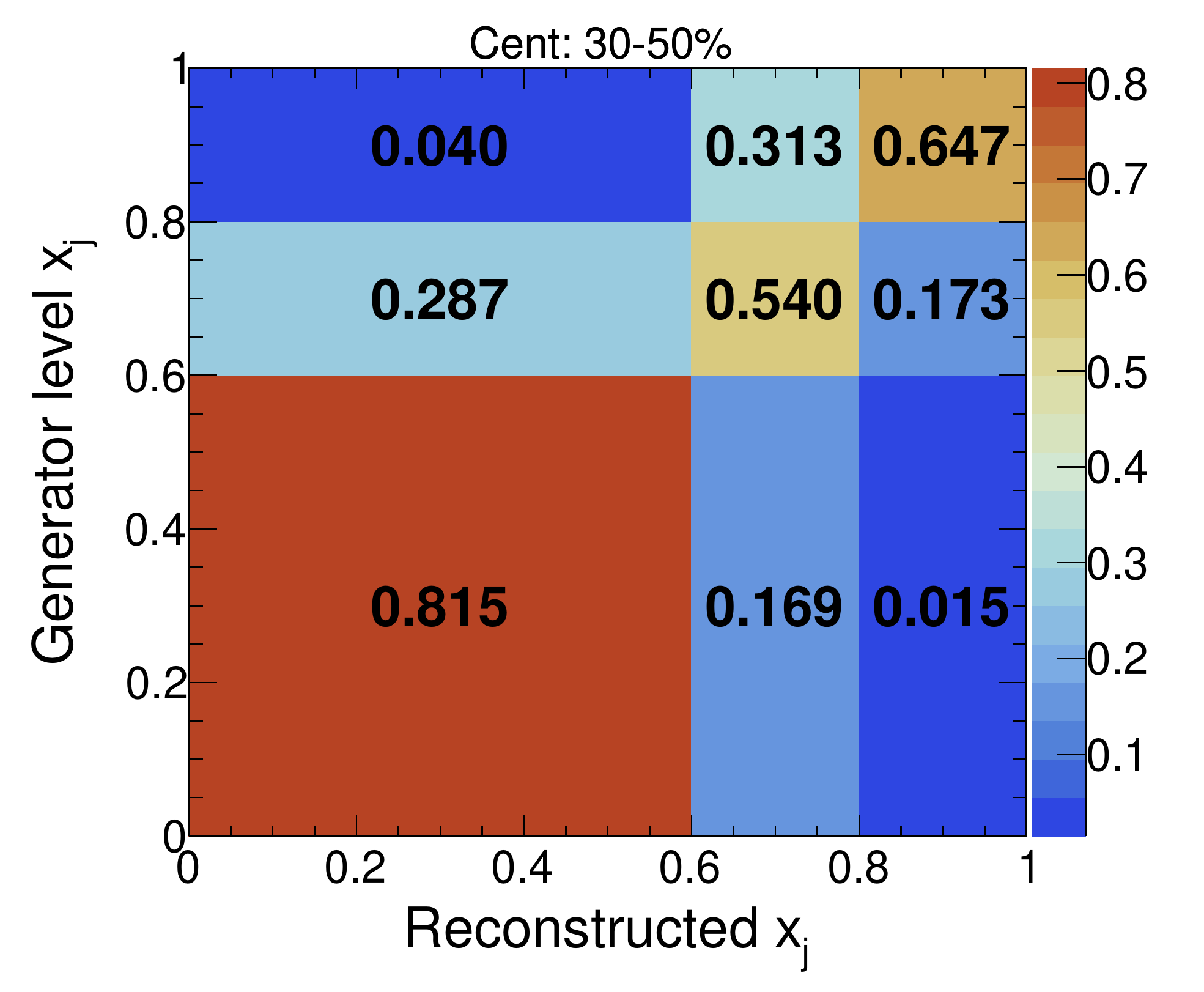} \\
    \includegraphics[width=0.32\linewidth]{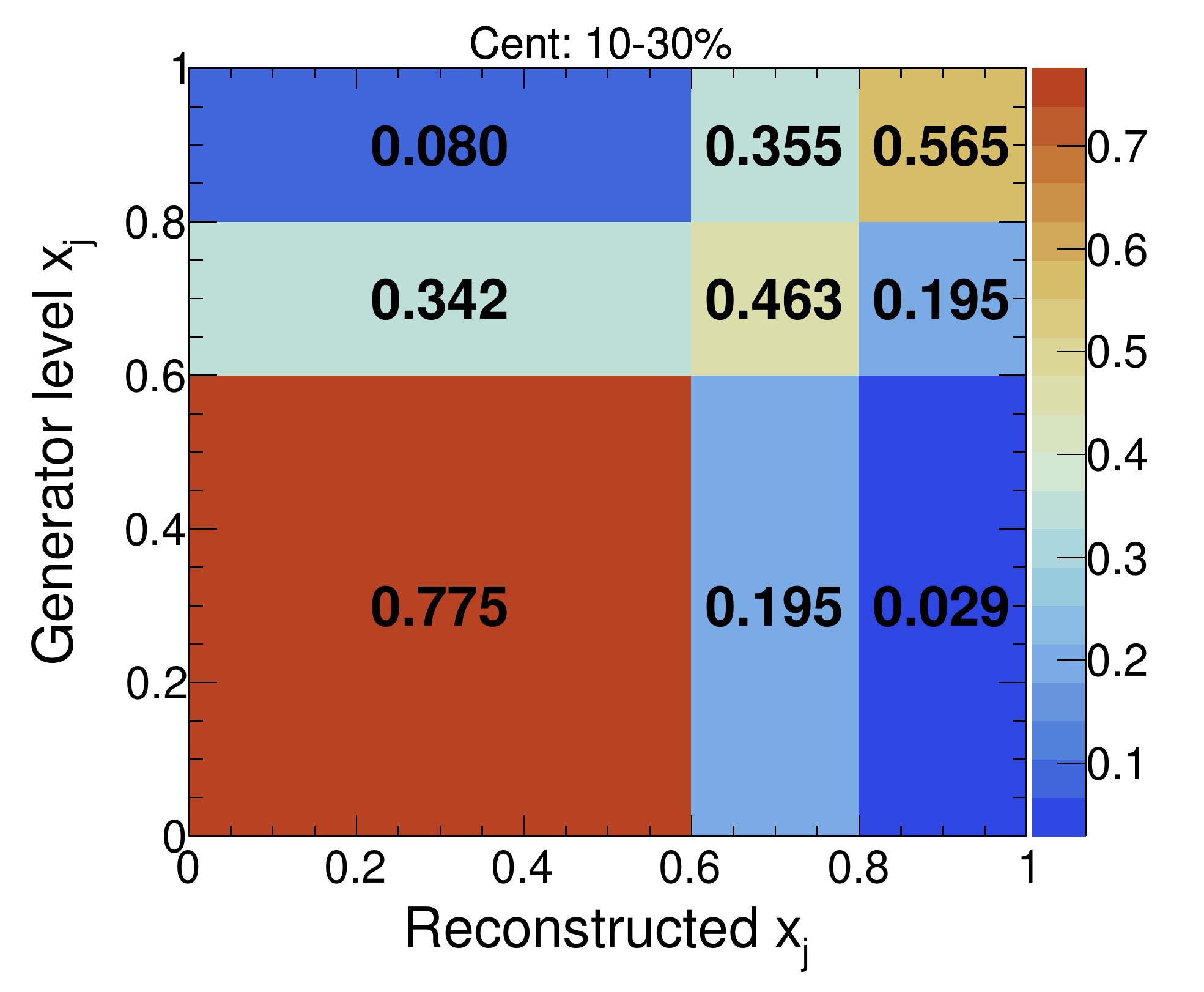}
    \includegraphics[width=0.32\linewidth]{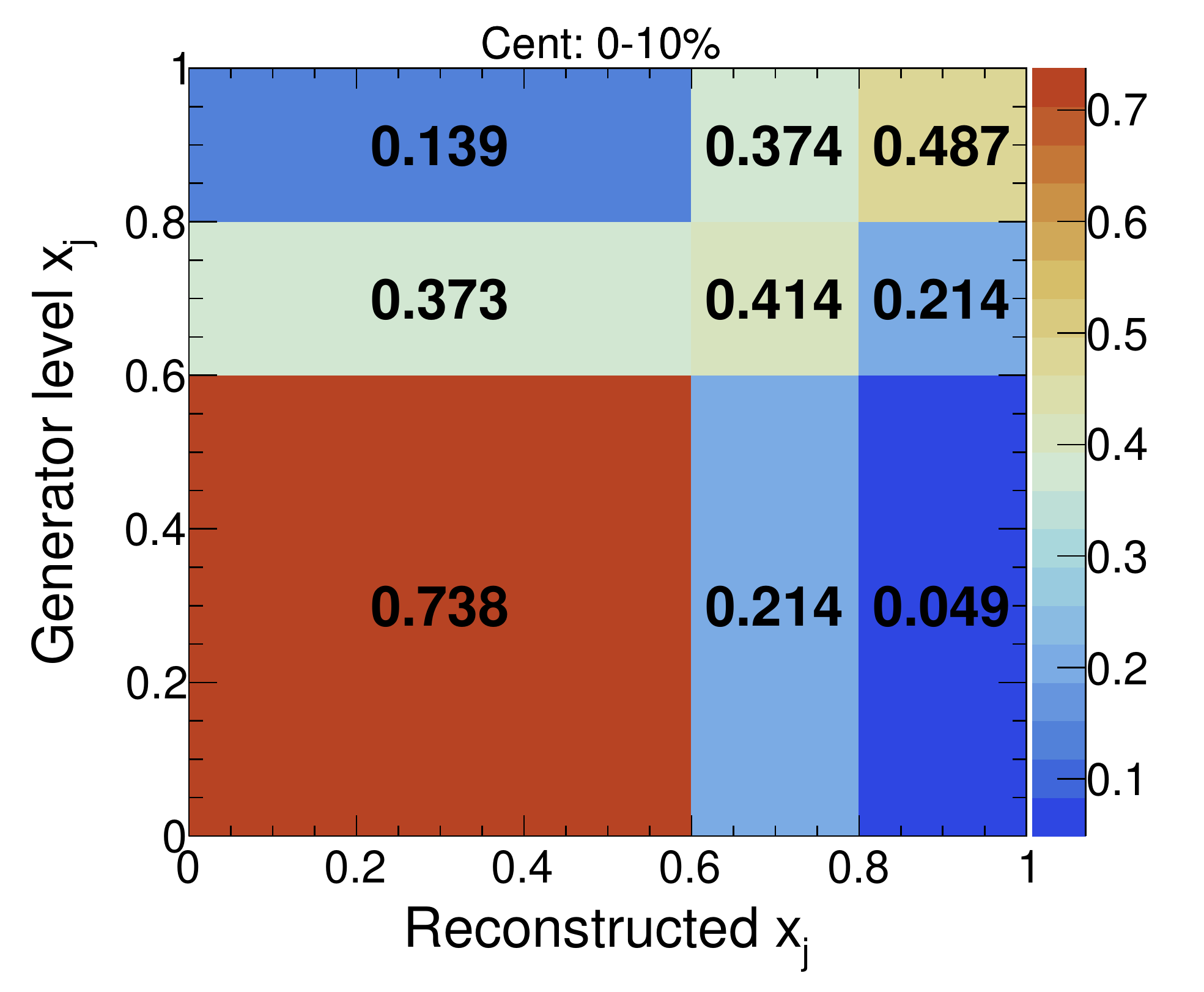}
    \caption{Generator-level vs.\ reconstructed \xj values in the analysis \xj bins. The plots show the probability to find a reconstructed \xj for a given generator level \xj. The \PYTHIA~8 simulation is shown in the upper-left plot while the most central {\PYTHIA}+\HYDJET is shown in the lower-right plot.}
  \label{fig:xjMatrixWideReverse}
  }
\end{figure}

Jet energy resolution effects may cause the \xj of the dijet system to migrate from one bin to another. The magnitude of this effect can be estimated from simulation. The plots illustrating \xj bin migrations in \PYTHIA~8 (\pp) and {\PYTHIA}+\HYDJET (PbPb) simulations are presented in Figs.~\ref{fig:xjMatrixWide} and~\ref{fig:xjMatrixWideReverse}. The plots in Fig.~\ref{fig:xjMatrixWide} give the probability for the generator level dijet to be in a specific \xj bin, given the \xj bin for the reconstructed dijet while those in Fig.~\ref{fig:xjMatrixWideReverse} give the probability for the reconstructed dijet to be in a specific \xj bin, given the \xj bin for the generator level dijet. For example, for the \PYTHIA~8 simulation in the leftmost plot of the top row in Fig.~\ref{fig:xjMatrixWide}, if the reconstructed dijet is in the bin $0 < \xj < 0.6$, the probability that the generator level dijet is also in this bin is 56.5\%. It can be seen from these plots that the \xj values in central heavy collisions are more strongly broadened than in \pp collisions in the simulation. In general, generator-level \xj values tend to be higher than the reconstructed ones, meaning that jet energy resolution effects cause dijets to be more unbalanced.

\cleardoublepage \section{The CMS Collaboration \label{app:collab}}\begin{sloppypar}\hyphenpenalty=5000\widowpenalty=500\clubpenalty=5000\vskip\cmsinstskip
\textbf{Yerevan Physics Institute, Yerevan, Armenia}\\*[0pt]
A.M.~Sirunyan$^{\textrm{\dag}}$, A.~Tumasyan
\vskip\cmsinstskip
\textbf{Institut f\"{u}r Hochenergiephysik, Wien, Austria}\\*[0pt]
W.~Adam, T.~Bergauer, M.~Dragicevic, J.~Er\"{o}, A.~Escalante~Del~Valle, R.~Fr\"{u}hwirth\cmsAuthorMark{1}, M.~Jeitler\cmsAuthorMark{1}, N.~Krammer, L.~Lechner, D.~Liko, T.~Madlener, I.~Mikulec, F.M.~Pitters, N.~Rad, J.~Schieck\cmsAuthorMark{1}, R.~Sch\"{o}fbeck, M.~Spanring, S.~Templ, W.~Waltenberger, C.-E.~Wulz\cmsAuthorMark{1}, M.~Zarucki
% ...
\vskip\cmsinstskip
\dag: Deceased\\
1:  Also at Vienna University of Technology, Vienna, Austria\\
% ...
\end{sloppypar}
%%% END EDITABLE REGION %%%
% skeleton_end
\end{document}